\documentclass[lettersize,journal]{IEEEtran}
\usepackage{amsmath,amsfonts}
\usepackage{algorithmic}
\usepackage{algorithm}
\usepackage{array}
\usepackage[caption=false,font=normalsize,labelfont=sf,textfont=sf]{subfig}
\usepackage{textcomp}
\usepackage{stfloats}
\usepackage{url}
\usepackage{verbatim}
\usepackage{graphicx}
\usepackage{cite}
\usepackage{amssymb}
\usepackage{float}
\usepackage{bbding}
\usepackage{lscape}
\usepackage{booktabs} 
\usepackage{multirow}
\usepackage{threeparttable}
\usepackage[dvipsnames]{xcolor}
\begin{document}

\title{HR-NVC++: Unified Ultra-High-Resolution Neural Video Coding with Confidence-Guided Spatial-Temporal Cues}


\author{Feng Ye, Lijiong Chen, Kai Zhang,~\IEEEmembership{Senior Member,~IEEE}, \\Li Zhang,~\IEEEmembership{Senior Member,~IEEE}, Chuanmin Jia,~\IEEEmembership{Member,~IEEE}
\thanks{Feng Ye, Lijiong Chen and Chuanmin Jia are with Wangxuan Institute of Computer Technology, Peking University, Beijing, China. }
\thanks{Kai Zhang and Li Zhang are with ByteDance Inc., San Diego, CA, USA.}
\thanks{Corresponding Author: Chuanmin Jia (cmjia@pku.edu.cn).}
}

\markboth{Preprint, Under Review}%
{Shell \MakeLowercase{\textit{et al.}}: A Sample Article Using IEEEtran.cls for IEEE Journals}


\maketitle

\begin{abstract}
Ultra-high-resolution neural video coding (NVC) remains challenging because temporal correspondence becomes increasingly unreliable as spatial resolution increases. Although recent neural B-frame codecs have improved bi-directional temporal prediction, complex motion and fine-grained structures still lead to unreliable temporal references, while spatial information of the current frame remains underexploited. 
In this work, we propose HR-NVC++, a unified ultra-high-resolution NVC approach that supports both fidelity-oriented and perceptual-quality-oriented compression with a shared spatial-temporal context. 
The context integrates bi-directional temporal cues with complementary spatial information: confidence-guided fusion adaptively balances the forward and backward temporal cues according to their estimated correspondence reliability, while a spatial complement branch enriches regions that temporal references fail to explain. The resulting context serves as a common basis for two compression modes. 
Built upon this shared context, the fidelity mode retains entropy-based latent coding for pixel-accurate compression, whereas the perceptual mode removes additional latent coding and performs reconstruction directly from the context.
This design demonstrates that a sufficiently informative spatial-temporal context already carries the dominant content required for perceptually favorable reconstruction, allowing different compression objectives to share the same contextual representation. 
Extensive experiments on UVG-4K and JVET Class A1/A2 show that HR-NVC++ achieves an average PSNR BD-rate saving of 13.01\% over the H.266/VVC reference model VTM-13.2 under the random access configuration, while delivering substantially better perceptual quality and temporal consistency than existing neural video codecs.

\end{abstract}

  

\section{Introduction}
\label{sec:intro}
\begin{figure}[tb]
  \centering
  \footnotesize
  
    \includegraphics[width=\linewidth]{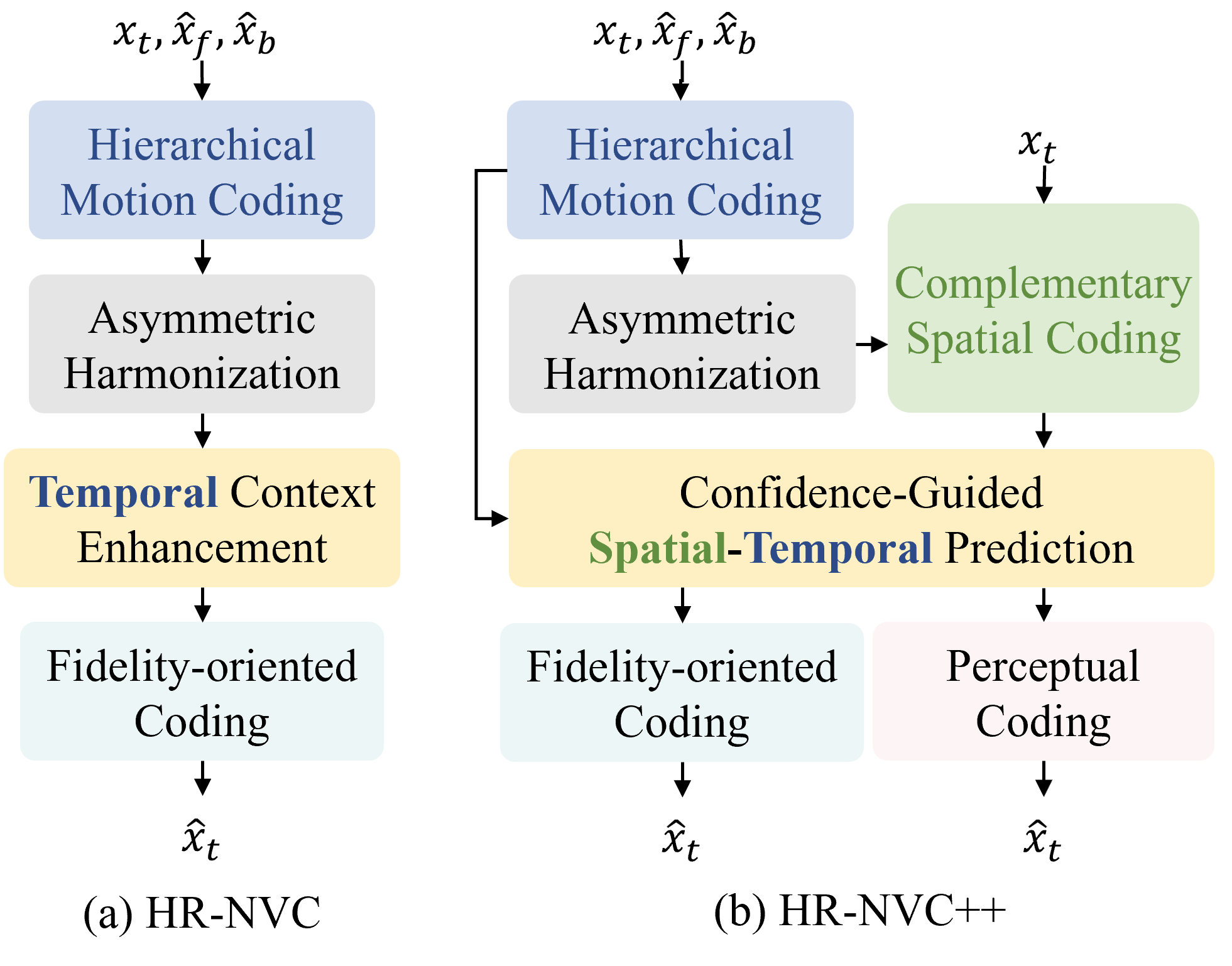}
  \caption{Comparison between (a) our prior model HR-NVC~\cite{ye2026high} and (b) HR-NVC++. HR-NVC enhances the temporal context for fidelity-oriented coding. HR-NVC++ further introduces complementary spatial coding and confidence-guided spatial-temporal prediction for both fidelity-oriented and perceptual coding. Here, $x_t$ denotes the current frame, $\hat{x}_f$/$\hat{x}_b$ denote the forward/backward reconstructed reference frames, and $\hat{x}_t$ denotes the reconstructed current frame.}
  \label{fig:comparison}
\end{figure}

\begin{figure*}[tb]
  \centering
  \footnotesize
  
    \includegraphics[width=\linewidth]{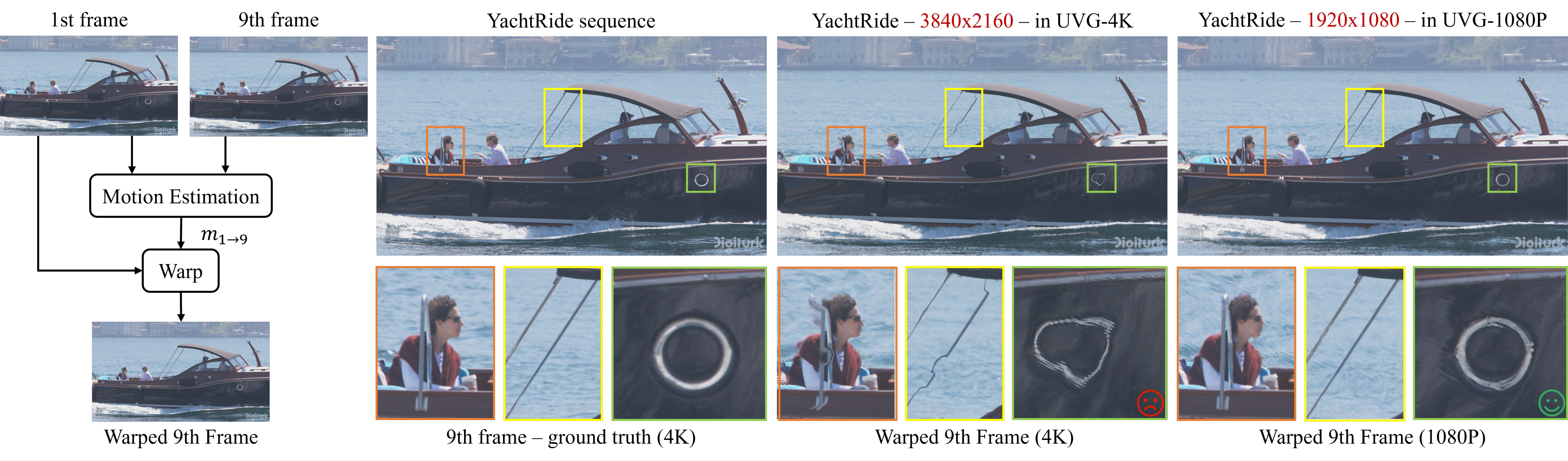}
  \caption{Resolution-dependent degradation of temporal warping on the YachtRide sequence. The warped 9th frame is predicted by warping the 1st frame with the estimated motion $m_{1\rightarrow 9}$, where \textbf{motion estimation and warping are performed at 4K or at 1080P, respectively}. The enlarged regions compare the 4K ground truth with the two warped predictions: the 4K prediction exhibits noticeable misalignment and distortion in fine structures, whereas the 1080P prediction preserves these structures faithfully, indicating that temporal correspondence becomes less reliable at ultra-high resolutions.}
  \label{fig:motivation}
\end{figure*}

Neural video coding becomes increasingly challenging as spatial resolution grows toward Ultra High Definition (UHD). UHD videos contain finer textures and more intricate structures, and motion spans a larger number of pixels, making accurate inter-frame correspondence increasingly difficult to establish. This challenge is particularly pronounced in hierarchical bi-directional prediction based video coding~\cite{Chen2022BCANFAB,sheng2024bi,deepye2024,jiang2025biecvc,liu2025brhvc,gao2025maskcrtb,ye2026high,jiang2026bicrvc}, where the current frame is often predicted from temporally distant reference frames. To alleviate this challenge, HR-NVC~\cite{ye2026high} introduces hierarchical motion representation and asymmetric context harmonization, improving the effectiveness of temporal prediction under complex motion and occlusion (Fig.~\ref{fig:comparison}(a)). Nevertheless, temporal references alone still fail in fine-grained regions at ultra-high resolutions. As illustrated in Fig.~\ref{fig:motivation}, temporal prediction that preserves local structures at 1080P can exhibit substantial misalignment and detail degradation at 4K. Such resolution-dependent degradation highlights a fundamental limitation of relying predominantly on temporal prediction for UHD video coding.

\IEEEpubidadjcol

However, unreliable temporal correspondence does not imply that the current frame is inherently unpredictable. Temporal references still provide valuable information about inter-frame structure and motion, while spatial information from the current frame itself can supply complementary evidence for regions that temporal prediction explains poorly, such as local edges, textures, and fine-grained details. 
A more effective contextual representation should therefore preserve reliable temporal evidence and incorporate spatial information where temporal correspondence is insufficient.

Beyond enriching the contextual representation, a related question is how such a context should support different compression objectives. Most NVCs adopt a rate-distortion formulation in which the current frame is encoded into a frame-specific latent conditioned on the available context. This latent is quantized and entropy-coded for transmission, providing additional information required for pixel-accurate reconstruction~\cite{9072487,li2021deep,rippel2021elf,Ho2022CANFVCCA,Shi2022AlphaVCHA,Li_2022,9880063,jia2023mpai,li2023neural,Kim2023NeuralVC,Alexandre2023HierarchicalBV,li2024neural,sheng2024spatial,hu2025evnvc}. 
Such explicit latent coding is essential for fidelity-oriented compression. Perceptual-quality-oriented compression, however, prioritizes visual similarity rather than strict pixel-wise correspondence. 
This raises a more specific question: once a sufficiently informative spatial-temporal context has been constructed, can it serve as the sole informative input for perceptually favorable reconstruction? If so, the same context could be shared across compression objectives.

To address these challenges, we propose \textbf{HR-NVC++}, a unified ultra-high-resolution NVC approach for both fidelity-oriented and perceptual-quality-oriented compression (Fig.~\ref{fig:comparison}(b)). HR-NVC++ constructs a shared spatial-temporal context by integrating bi-directional temporal cues and complementary spatial information. Specifically, confidence-guided fusion adaptively balances the forward and backward temporal cues according to their estimated reliability, while a spatial complement branch enriches regions that remain insufficiently represented by temporal references. The resulting context provides a common basis for compression under different quality objectives.

Based on this shared context, HR-NVC++ supports two compression modes. The fidelity mode retains entropy-based latent coding: the current-frame latent is quantized and entropy-coded to preserve additional frame-specific information for pixel-accurate reconstruction. 
In contrast, the perceptual decoder uses the shared context as its sole informative input, without requiring additional entropy-coded information.
This formulation allows us to examine whether the contextual representation itself can provide the dominant content information required for perceptual reconstruction, while allowing the compression objective to change without modifying the context-construction process.


As summarized in Fig.~\ref{fig:comparison}, this article extends our previous work, HR-NVC~\cite{ye2026high}, in the following four aspects:
\begin{itemize}
    \item \textbf{Complementary spatial coding.} HR-NVC++ introduces a spatial complement branch to provide local structural evidence in regions where bi-directional temporal references are unreliable.

    \item \textbf{Confidence-guided spatial-temporal fusion.} With the spatial complement as guidance, HR-NVC++ performs group-based refinement of bi-directional motions and confidences, followed by adaptive fusion to construct the spatial-temporal context.

    \item \textbf{Unified coding objectives.} Both fidelity-oriented and perceptual-quality-oriented modes share the same spatial-temporal context. The fidelity mode retains entropy-coded latent coding, whereas the perceptual mode reconstructs directly from the context.
    
    \item \textbf{Two-stage perceptual fine-tuning.} The pretrained fidelity-oriented model is fine-tuned in two stages toward perceptual-quality-oriented compression, with the context construction module kept fixed, enabling perceptual reconstruction solely from the shared context at inference.
    
\end{itemize}

The main contributions are summarized as follows:

\begin{itemize}
    \item We propose HR-NVC++, a unified ultra-high-resolution NVC approach that supports both fidelity-oriented and perceptual-quality-oriented compression based on the same spatial-temporal context. 
    
    \item We propose a confidence-guided spatial-temporal context for UHD video coding, in which bi-directional temporal cues are fused according to their reliability and enriched by complementary spatial information.

    \item We introduce a two-stage perceptual fine-tuning strategy that enables perceptual reconstruction with the shared context as the sole informative input at inference.
    
    \item  Extensive experimental results show that, compared with the H.266/VVC reference model VTM-13.2 under random access configuration, HR-NVC++ achieves average PSNR BD-rate savings of 13.01\% on 4K benchmarks, together with consistent gains in perceptual quality and temporal consistency.
\end{itemize}
      
\section{Related Work}
\label{sec:related}

\subsection{Neural P-frame Based Video Coding}
\label{sec:related_video}
End-to-end neural video compression was established under the motion-compensation and residual-coding paradigm. DVC~\cite{lu2019dvc} jointly optimized motion estimation, motion compression, and residual coding within an end-to-end rate-distortion framework, demonstrating the feasibility of a fully learned video codec. FVC~\cite{hu2021fvc} further moved these operations into the feature space, enabling more effective exploitation of spatial-temporal dependencies. A major shift was then introduced by DCVC~\cite{li2021deep}, which replaced explicit residual coding with conditional coding: previously decoded frames are transformed into temporal contexts that condition both the reconstruction and the entropy modeling of the current frame. This conditional paradigm has since become dominant in neural video compression.

Subsequent advances under this paradigm can be grouped into three directions. First, the contextual representation itself has been continuously enriched, from temporal context mining~\cite{sheng2021temporal} and diverse context learning~\cite{li2023neural} to feature modulation with periodic refreshing~\cite{li2024neural} and non-local correlation modeling across multiple reference frames~\cite{jiang2025ecvc}. Second, the utilization of the context has been improved through more powerful conditional distribution modeling, including hybrid spatial-temporal entropy models~\cite{Li_2022}, context-guided entropy design~\cite{cgt2025}, and content-adaptive hierarchical hyperpriors~\cite{liao2026contentadaptive}. Third, recent efforts have pushed the paradigm toward practical deployment, through hierarchical quality structures~\cite{ehvc2025}, real-time oriented design~\cite{dcvcrt2025}, and unified or ultrafast architectures~\cite{xiang2026uiic,li2026dcvcuf}.


\subsection{Neural Bi-directional Prediction Based Video Coding}
\label{sec:related_bframe}
Compared with P-frame coding, neural B-frame coding requires effective exploitation of information from both temporal directions within an efficient hierarchical coding structure. Early attempts formulated bi-directional prediction as frame interpolation~\cite{wu2018video} or by directly extending P-frame codecs~\cite{pourreza2021extending}, demonstrating the potential of bi-directional references without fully exploiting them for compression.

Subsequent studies have developed learned B-frame coding along three lines. The first line focuses on the coding structure: B-CANF~\cite{Chen2022BCANFAB} introduced conditional augmented normalizing flows for adaptive bi-directional prediction, ~\cite{Alexandre2023HierarchicalBV} eliminated explicit motion coding with a two-layer CANF architecture, ~\cite{Kim2023NeuralVC} introduced temporal layer-adaptive optimization for hierarchical B-frame coding, and MaskCRT-B~\cite{gao2025maskcrtb} employed a masked conditional residual transformer. The second line extends the conditional coding paradigm to bi-directional prediction: DCVC-B~\cite{sheng2024bi} introduced bi-directional temporal context propagation and entropy modeling, and the motion and prediction representations were further improved by a scaled hierarchical bi-directional motion model~\cite{deepye2024} and a spatial-temporal decomposition of inter prediction~\cite{sheng2024spatial}. The third line coordinates the asymmetric reliability of the two references: BiECVC~\cite{jiang2025biecvc} combined diversified local and non-local contexts with adaptive gating, and BRHVC~\cite{liu2025brhvc} performed bi-directional motion convergence and contextual harmonization.

Our previous work, HR-NVC~\cite{ye2026high}, revisited high-resolution B-frame coding from the perspective of reference information modeling, organizing bi-directional reference information into representation, translation, and harmonization. Nevertheless, the context in existing bi-directional codecs is still constructed from temporal references alone. HR-NVC++ therefore introduces complementary spatial information from the current frame into the context construction, and coordinates temporal and spatial evidence through confidence-guided spatial-temporal fusion, so that regions poorly covered by temporal references can be compensated at a low coding cost. Moreover, the resulting spatial-temporal context is shared by both fidelity-oriented and perceptual reconstruction, which distinguishes HR-NVC++ from existing B-frame codecs designed for a single fidelity objective.

\subsection{Perceptual-Quality-Oriented Neural Video Coding}
\label{sec:related_perceptual}
Conventional learned video compression predominantly optimizes pixel-wise distortion, such as mean squared error (MSE). However, pixel-level fidelity does not always align with human perceptual preference, especially at low bitrates where faithful reconstruction becomes infeasible, motivating perceptual-quality-oriented neural video coding.

Existing approaches differ mainly in how visual realism is obtained. The first category introduces perceptual-oriented designs into the coding framework, as exemplified by PNVC-CR~\cite{Liang2026Perceptual} with color separation and rank chain. The second category relies on adversarial training, including recurrent conditional GANs for spatial-temporal consistency~\cite{yang2021perceptual}, GAN-based detail synthesis and propagation~\cite{mentzer2022gan}, and coding within a pre-trained GAN latent space~\cite{shukor2022gan}. The third category leverages diffusion priors, including one-step diffusion with temporal context guidance~\cite{ma2025diffvcosd}, single-step diffusion with semantic-temporal guidance~\cite{xue2026s2vc}, and video diffusion priors with sequence-level denoising~\cite{mao2026gnvc}. In contrast, HR-NVC++ derives its perceptual mode from a pretrained fidelity-oriented model through architectural modification and two-stage fine-tuning, achieving strong perceptual performance.

\section{Technical Details of HR-NVC++}
\label{sec:method}
\begin{figure*}[tb]
  \centering
    \includegraphics[width=\linewidth]{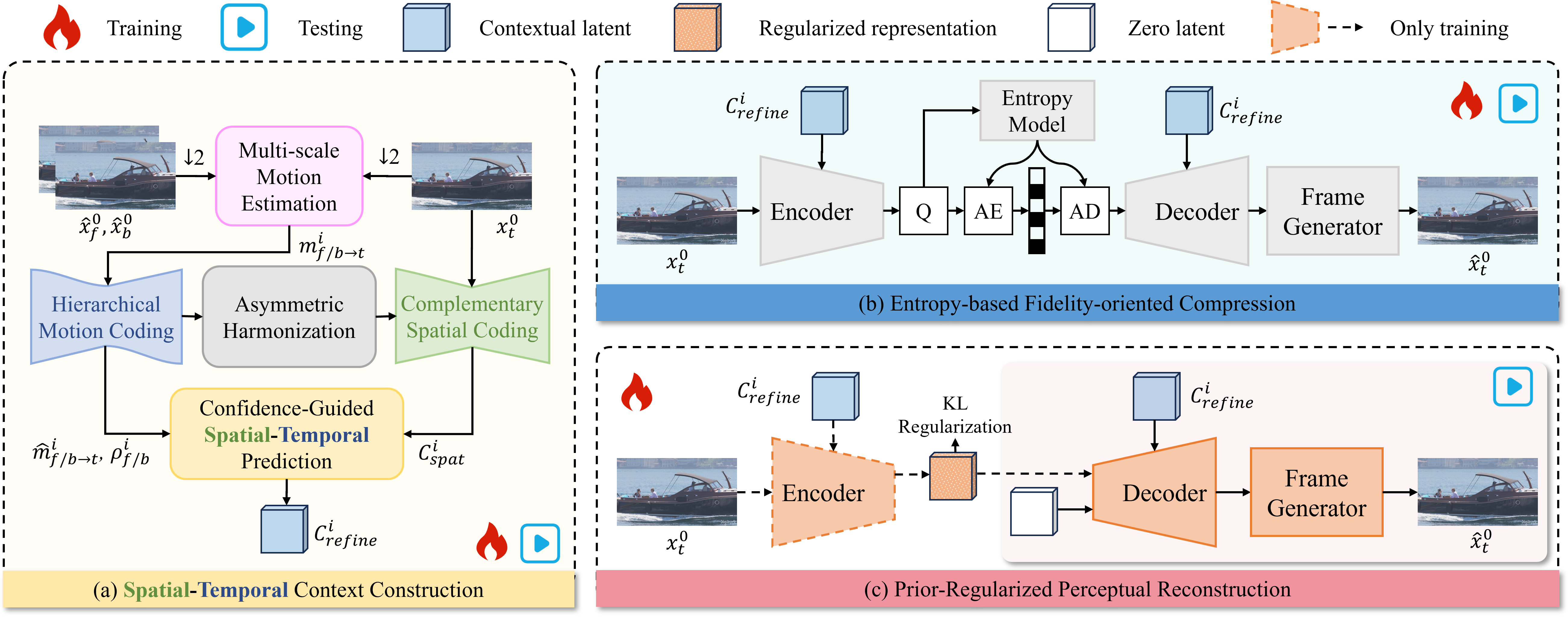}
    \caption{Overall architecture of HR-NVC++. 
    (a) The confidence-guided spatial-temporal context construction module integrates multi-scale motion, asymmetric harmonization, and complementary spatial information to construct the shared context. Based on this shared context, HR-NVC++ supports two modes: (b) \textbf{entropy-based fidelity-oriented compression}, which explicitly encodes the current-frame latent for pixel-accurate reconstruction, and (c) \textbf{prior-regularized perceptual reconstruction}, which uses the shared context as the sole informative input at inference.}
\label{fig:framework}
\end{figure*}


\subsection{Overview}

HR-NVC++ separates high-resolution video compression into two stages: shared spatial-temporal context generation and objective-specific compression or reconstruction, as illustrated in Fig.~\ref{fig:framework}. Given the current frame $x_t^0$ and its forward and backward reconstructed reference frames $\hat{x}_f^0$ and $\hat{x}_b^0$, the context generator constructs a shared representation by integrating bi-directional temporal information, correspondence confidence, and complementary spatial information from the current frame.

Specifically, hierarchical motion estimation establishes temporal correspondence across multiple resolutions, and confidence-guided fusion adaptively balances the forward and backward temporal contexts according to their estimated reliability. A spatial complement branch then provides additional evidence for regions that remain insufficiently represented by temporal references. By integrating this spatial information with the confidence-refined temporal features, HR-NVC++ produces a multi-scale shared spatial-temporal context, $\{C_{\mathrm{refine}}^i\}_{i=0}^{2}$, for subsequent compression or reconstruction. 

\begin{figure*}[tb]
  \centering
    \includegraphics[width=\linewidth]{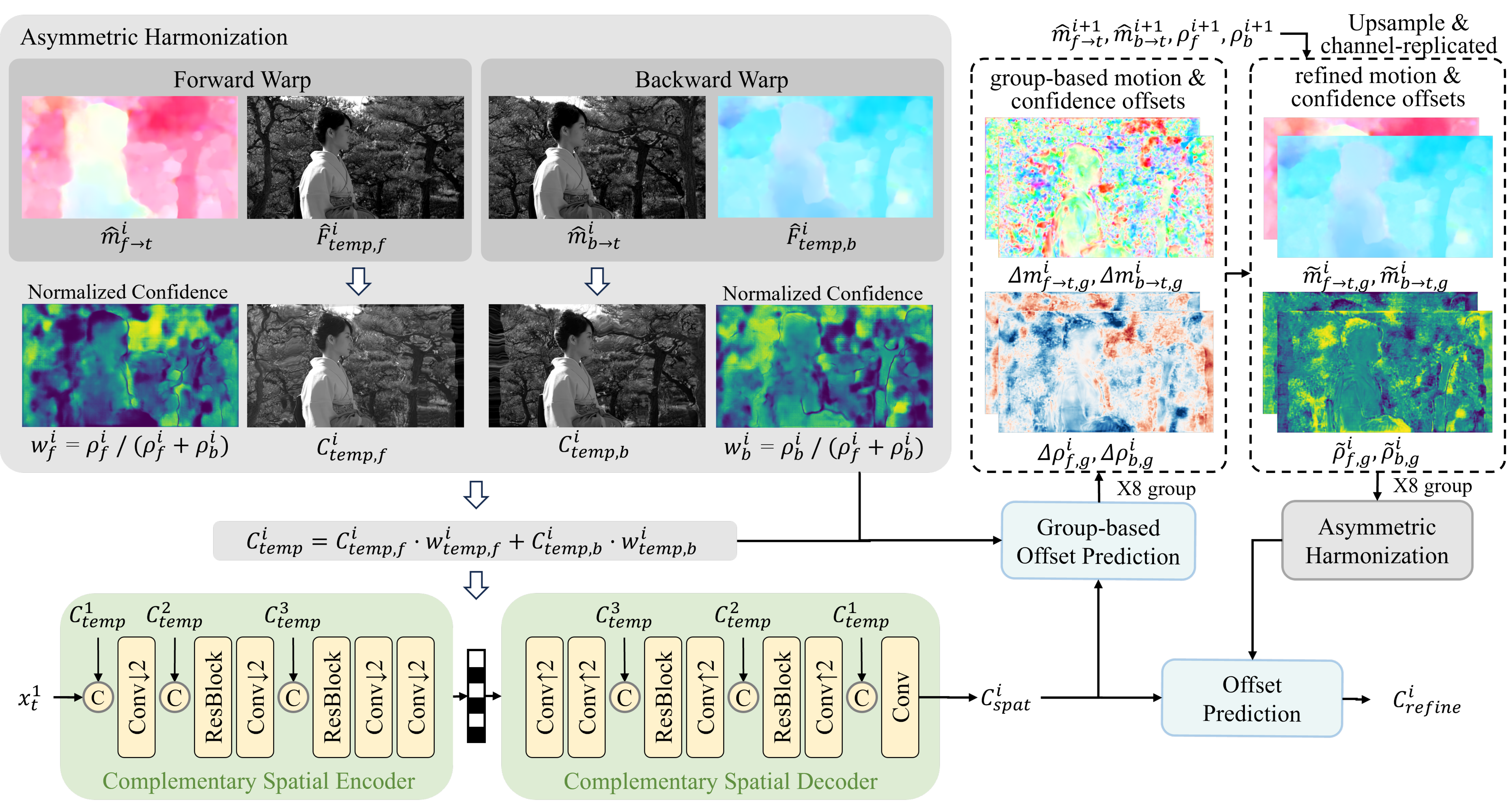}
  \caption{Illustration of the confidence-guided spatial-temporal context construction at scale $i$. The reference features $\hat{F}^i_{temp,f}$ and $\hat{F}^i_{temp,b}$ are warped to the current-frame coordinates with the reconstructed motion fields $\hat{m}^i_{f\rightarrow t}$ and $\hat{m}^i_{b\rightarrow t}$, and the two warped contexts are fused using the normalized confidence maps $w^i_f$ and $w^i_b$, yielding the fused temporal context $C^i_{temp}$. The complementary spatial codec encodes the half-resolution current frame $x^1_t$ conditioned on the temporal contexts, producing the spatial complement $C^i_{spat}$. The spatial complement, fused temporal context and the upsampled and channel-replicated motion fields and confidence maps from the coarser scale $i{+}1$, are then used to predict the group-based motion and confidence offsets, yielding the refined motions and confidence maps. The refined temporal representation is finally enhanced with the spatial complement, producing the shared context $C^i_{refine}$.}
  \label{fig:confidence_map}
\end{figure*}

Based on the shared context, HR-NVC++ supports two modes: \textbf{fidelity-oriented compression} and \textbf{perceptual reconstruction}. In the fidelity mode, the current frame is encoded conditioned on the shared context, and the resulting latent is quantized and entropy-coded to preserve additional frame-specific information for pixel-accurate reconstruction. In contrast, the perceptual mode does not transmit this frame-dependent latent: at inference, the decoder reconstructs the current frame directly from the shared context with a fixed zero input. Thus, both modes share the same contextual representation but differ in their reconstruction formulations.

\subsection{Shared Spatial-Temporal Context Construction}
\label{sec:context}
The shared spatial-temporal context is constructed through four steps: hierarchical bi-directional motion representation, confidence-guided temporal context fusion, complementary spatial coding, and confidence-guided spatial-temporal fusion. The first two steps establish a temporal representation together with its correspondence reliability. The spatial complement then provides additional structural information from the current frame, which is further integrated into the temporal representation through group-wise motion and confidence refinement. The resulting multi-scale contexts are shared by the fidelity-oriented compression and perceptual reconstruction branches.

\subsubsection{Hierarchical Bi-directional Motion Representation}
\label{sec:motion}
Accurate motion estimation at the original resolution can be challenging in the presence of large displacements and fine-grained structures. We therefore adopt a coarse-to-fine strategy that first estimates dominant motion at a coarse resolution and progressively refines the motion fields at finer scales.

Let $x_t^i$, $\hat{x}_f^i$, and $\hat{x}_b^i$ denote the current frame and its forward and backward reconstructed reference frames at scale $i$, where $i=0,1,2,3$ correspond to the original, $1/2$, $1/4$, and $1/8$ spatial resolutions, respectively. At the coarsest scale, the initial bi-directional motion fields are estimated as
\begin{equation}
m_{f\rightarrow t}^{3}=\mathcal{F}\left(x_t^3,\hat{x}_f^3\right),\ 
m_{b\rightarrow t}^{3}=\mathcal{F}\left(x_t^3,\hat{x}_b^3\right).
\label{eq:coarse_motion}
\end{equation}
where $\mathcal{F}(\cdot)$ denotes the motion estimation network. The estimated motion fields are then propagated to finer scales and progressively refined:
\begin{equation}
\begin{aligned}
m_{f\rightarrow t}^{i}&=\mathcal{F}\left(x_t^i,\hat{x}_f^i,U(m_{f\rightarrow t}^{i+1})\right),\\ 
m_{b\rightarrow t}^{i}&=\mathcal{F}\left(x_t^i,\hat{x}_b^i,U(m_{b\rightarrow t}^{i+1})\right),
\label{eq:fine_motion}
\end{aligned}
\end{equation}
where $U(\cdot)$ denotes upsampling. Here, the motion estimated at the coarser scale provides an initialization for refining local correspondences at the next finer scale, allowing the model to stably capture large displacements while progressively improving alignment accuracy.

Following HR-NVC~\cite{ye2026high}, the multi-scale bi-directional motion fields are jointly compressed into a compact representation. After quantization and decoding, the reconstructed motion fields and their associated confidence maps are obtained as:
\begin{equation}
\left\{\hat{m}_{f\rightarrow t}^{i},\hat{m}_{b\rightarrow t}^{i},\rho_f^i,\rho_b^i\right\}_{i=1}^{3}.
\label{eq:motion_decode}
\end{equation}
Note that the motion at the original resolution is not estimated or coded; only the motion fields at scales $i=1,2,3$ (half resolution and below) are transmitted, which substantially reduces both the bitrate and the computational cost of motion modeling. The confidence maps $\rho_f^i$ and $\rho_b^i$ characterize the estimated reliability of forward and backward temporal correspondences, respectively. These maps are subsequently used to adapt the relative contributions of the two temporal references.

\subsubsection{Confidence-Guided Temporal Context Fusion}
\label{sec:temporal_context}
As illustrated in Fig.~\ref{fig:confidence_map}, the reconstructed motion fields are used to align the forward and backward reference features with the current frame. The resulting temporal contexts are:
\begin{equation}
\begin{aligned}
C^i_{temp,f} = \mathcal{W}\left(\hat{F}^i_{temp,f}, \hat{m}^i_{f\rightarrow t}\right), \\
C^i_{temp,b} = \mathcal{W}\left(\hat{F}^i_{temp,b}, \hat{m}^i_{b\rightarrow t}\right).
\label{eq:warp_context}
\end{aligned}
\end{equation}
Here, $\hat{F}^i_{temp,f}$ and $\hat{F}^i_{temp,b}$ denote the temporal reference features at scale $i$ extracted from $\hat{x}_f$ and $\hat{x}_b$, respectively, and $\mathcal{W}(\cdot)$ denotes feature warping using the corresponding reconstructed motion field.

The reliability of the two temporal references varies across spatial locations due to occlusion and complex motion. To account for this asymmetry, we normalize the confidence maps to obtain spatially adaptive fusion weights:
\begin{equation}
w^i_{f} = \frac{\rho^i_f}{\rho^i_f + \rho^i_b}, \
w^i_{b} = \frac{\rho^i_b}{\rho^i_f + \rho^i_b}.
\label{eq:temporal_weights}
\end{equation}
The bi-directional temporal context is then obtained by element-wise weighted fusion:
\begin{equation}
C^i_{temp} = w^i_{f} \odot C^i_{temp,f} + w^i_{b} \odot C^i_{temp,b}.
\label{eq:temporal_context}
\end{equation}
This formulation allows the model to assign larger contributions to the more reliable temporal reference at each location, providing a spatially adaptive temporal representation for subsequent context refinement.

\subsubsection{Complementary Spatial Coding}
\label{sec:spatial_complement}
Even after confidence-guided bi-directional fusion, temporal references may not adequately represent all local structures in a high-resolution frame. We therefore introduce a spatial complement branch to provide additional structural information from the current frame.

Rather than encoding the current frame independently of the temporal context, the spatial branch is conditioned on the available temporal contexts. Specifically, the current frame is first downsampled to half resolution, and the spatial complement is obtained as
\begin{equation}
\left\{ C^i_{spat} \right\}_{i=0}^{2} = D_c\left( Q\left( E_c\left( x^1_t, \left\{ C^i_{temp} \right\}_{i=1}^{3} \right) \right) \right).
\label{eq:spatial_multiscale}
\end{equation}
Here, $E_c$, $Q$, and $D_c$ denote the complementary spatial encoder, quantization operation, and decoder, respectively. Conditioning the spatial representation on the temporal context enables the branch to provide complementary structural evidence for regions that temporal prediction may not adequately explain, including local edges, textures, and fine-grained details. The resulting multi-scale spatial features are then used to refine the temporal representation.

\subsubsection{Confidence-Guided Spatial-Temporal Fusion}
\label{sec:st_fusion}
The confidence-fused temporal context and the spatial complement are jointly exploited to form the spatial-temporal context. At each scale, a refinement module takes the spatial complement, temporal reference features, and upsampled motion and confidence maps as input, and predicts group-specific residual corrections:
\begin{equation}
\begin{aligned}
&\left\{ \Delta m^i_{f\rightarrow t,g}, \Delta m^i_{b\rightarrow t,g}, \Delta \rho^i_{f,g}, \Delta \rho^i_{b,g} \right\}_{i=0}^{2} = \phi^i( C^i_{spat}, \\
&\hat{F}^{i}_{temp,f}, \hat{F}^{i}_{temp,b}, U(\hat{m}^{i+1}_{f\rightarrow t}), U(\hat{m}^{i+1}_{b\rightarrow t}), U(\rho^{i+1}_f), U(\rho^{i+1}_b) ),
\label{eq:group_refinement}
\end{aligned}
\end{equation}
where $\phi^i(\cdot)$ predicts the group-wise motion and confidence corrections.

The features are organized into $G=8$ channel groups. For each group, the corresponding motion fields and confidence maps are refined using the predicted residuals:
\begin{equation}
\begin{aligned}
\tilde{m}^i_{f\rightarrow t,g} &= R\left(U(\hat{m}^{i+1}_{f\rightarrow t})\right) + \Delta m^i_{f\rightarrow t,g}, \\
\tilde{m}^i_{b\rightarrow t,g} &= R\left(U(\hat{m}^{i+1}_{b\rightarrow t})\right) + \Delta m^i_{b\rightarrow t,g}, \\
\tilde{\rho}^i_{f,g} &= R\left(U(\rho^{i+1}_f)\right) + \Delta \rho^i_{f,g}, \\
\tilde{\rho}^i_{b,g} &= R\left(U(\rho^{i+1}_b)\right) + \Delta \rho^i_{b,g}.
\end{aligned}
\end{equation}
Here, $R(\cdot)$ denotes replication of the motion fields and confidence maps along the feature-channel dimension to provide corresponding representations for each channel group.

The refined group-wise motion fields are then used to re-warp the reference features, producing the group-wise temporal contexts:
\begin{equation}
\begin{aligned}
C^i_{f,g} &= \mathcal{W}(\hat{F}^i_{temp,f}, \tilde{m}^i_{f\rightarrow t,g}),\\
C^i_{b,g} &= \mathcal{W}(\hat{F}^i_{temp,b}, \tilde{m}^i_{b\rightarrow t,g}).
\end{aligned}
\end{equation}
The refined confidence maps are normalized to obtain the group-specific fusion weights $(\tilde{w}^i_{f,g}, \tilde{w}^i_{b,g})$. The temporal context for the $g$-th channel group is then computed as:
\begin{equation}
C^i_g = \tilde{w}^i_{f,g} \odot C^i_{f,g} + \tilde{w}^i_{b,g} \odot C^i_{b,g}.
\end{equation}
The group-wise temporal contexts are concatenated to form the fused temporal context:
$C^i = \operatorname{Concat}_{g=1}^{G}(C_g^i)$.

Finally, the group-wise temporal representation is further enhanced with the spatial complement:
\begin{equation}
C^i_{refine} = C^i + \psi^i\left( C^i, C^i_{spat} \right),
\end{equation}
where $\psi^i(\cdot)$ denotes the spatial-temporal refinement operation. The resulting multi-scale contexts $\{C^i_{refine}\}_{i=0}^{2}$ form the shared spatial-temporal representation used by both fidelity-oriented compression and perceptual reconstruction.

\begin{figure*}[tb]
  \centering
    \includegraphics[width=0.95\linewidth]{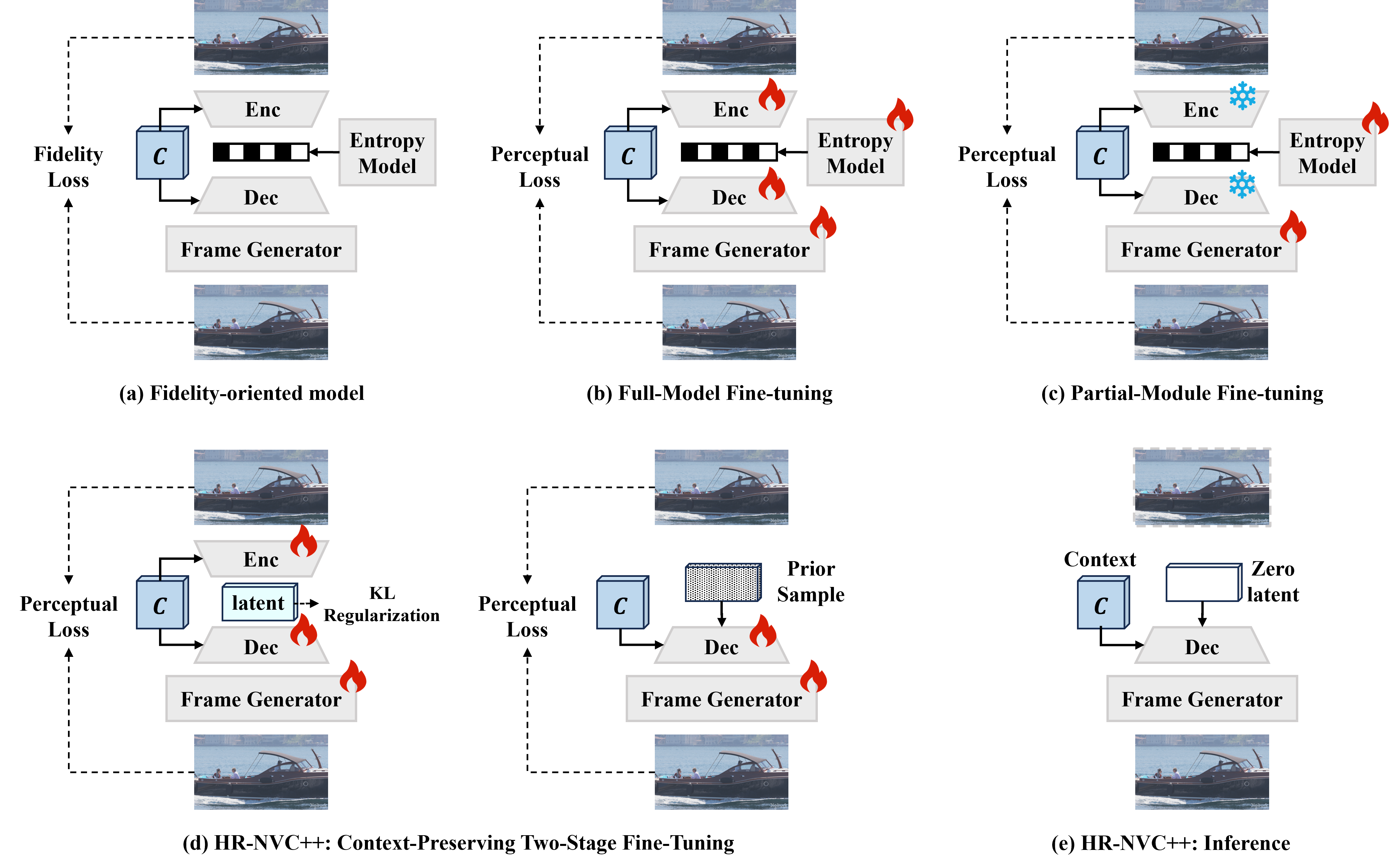}
  \caption{Comparison of training strategies for perceptual compression from a pretrained fidelity-oriented NVC. (a) The pretrained fidelity-oriented model. (b) Full-model and (c) partial-module fine-tuning without architectural modification. (d) The proposed context-preserving two-stage fine-tuning. Stage 1 regularizes the frame-dependent latent toward a standard Gaussian prior and fine-tunes the model using the perceptual objective. Stage 2 removes the encoder and further fine-tunes the reconstruction modules using prior samples. (e) At inference, a fixed zero-valued latent is fed to the decoder, with the shared context as the sole informative input. The context construction modules remain fixed during both training stages.}
  \label{fig:adaptation}
\end{figure*}

\subsection{Fidelity-Oriented Reconstruction}
\label{sec:fidelity}

For fidelity-oriented reconstruction, HR-NVC++ retains the conventional rate-distortion coding framework. The shared spatial-temporal context provides complementary spatial and temporal information, while the explicitly coded latent preserves additional frame-specific information required for pixel-accurate reconstruction.

Given the current frame $x_t$ and the shared spatial-temporal context $C_{\mathrm{refine}}^i$, the fidelity encoder $E_F(\cdot)$ produces a latent representation:
\begin{equation}
y_t^F = E_F(x_t,\left\{C_{\mathrm{refine}}^i\right\}_{i=0}^{2}).
\label{eq:fidelity-encoding}
\end{equation}
The latent is quantized and entropy-coded for transmission: 
\begin{equation}
\hat{y}_t^F = Q(y_t^F).
\label{eq:fidelity-quantization}
\end{equation}

At the fidelity decoder $D_F(\cdot)$, the reconstructed frame is obtained by combining the decoded latent with the shared context:
\begin{equation}
\hat{x}_t^F = D_F(\hat{y}_t^F,\left\{C_{\mathrm{refine}}^i\right\}_{i=0}^{2}).
\label{eq:fidelity-decoding}
\end{equation}

The fidelity branch is optimized using rate-distortion objective:
\begin{equation}
\mathcal{L}_F = R_t + \lambda_F D_{MSE}(x_t, \hat{x}^F_t),
\label{eq:fidelity-loss}
\end{equation}
where $R_t$ denotes the total estimated coding rate of all transmitted components, including the motion fields, the spatial complement, and the frame latent, and $\lambda_F$ controls the trade-off between bitrate and reconstruction fidelity.

\section{Prior-Regularized Perceptual Optimization}
\label{sec:perceptual}
The shared spatial-temporal context provides rich information for reconstructing the current frame. To explore its potential for perceptual reconstruction, we fine-tune the pretrained fidelity-oriented model while keeping the shared context generation module fixed.

Fig.~\ref{fig:adaptation} compares this strategy with existing fine-tuning approaches. In a fidelity-oriented model (Fig.~\ref{fig:adaptation}(a)), the current frame is encoded into a latent conditioned on the context, which is quantized and entropy-coded for transmission. To optimize such a model for perceptual reconstruction, existing approaches either fine-tune the entire model (Fig.~\ref{fig:adaptation}(b)) or only partial modules (Fig.~\ref{fig:adaptation}(c)). Both strategies keep the model architecture unchanged, and separate sets of weights have to be stored for the fidelity-oriented and perceptual-quality-oriented objectives. In contrast, HR-NVC++ adopts a context-preserving reconstruction strategy (Fig.~\ref{fig:adaptation}(d) and (e)): during inference, the decoder uses a fixed zero latent, with the shared context as the sole informative input. In this way, not only the contextual representation but also the weights of the context generation module are shared across the two compression objectives.

The fine-tuning process consists of two stages: the first stage converts the pretrained latent encoder into a conditional Gaussian parameterization regularized toward the standard Gaussian prior (Sections~\ref{sec:4-1} and ~\ref{sec:4-2}), and the second stage removes the encoder and fine-tunes the perceptual reconstruction modules with prior samples, so that inference requires only a fixed zero input (Section~\ref{sec:4-3}). This two-stage adaptation provides a smooth transition from the informative encoder output to the uninformative prior, avoiding the difficulty of directly optimizing the decoder with prior samples from scratch. 

\subsection{Conditional Gaussian Parameterization}
\label{sec:4-1}

We initialize the perceptual reconstruction branch from the pretrained fidelity-oriented model. During the first optimization stage, the latent encoder is retained, but its output is no longer quantized and entropy-coded for transmission. Instead, we parameterize the latent representation as a conditional Gaussian distribution.

Specifically, the original final layer is retained to predict the mean, and a duplicated copy of this layer is introduced to predict the log-variance:
\begin{equation}
\mu_t, \log \mathrm{var}_t = E_{P}\left(x_t, \{C^i_{refine}\}_{i=0}^{2}\right).
\label{eq:perceptual_mean_logvar}
\end{equation}
The duplicated layer is initialized with zero weights and bias, so that its initial output is zero; the conditional distribution therefore starts with unit variance and is already close to the standard Gaussian prior at the beginning of training.

The resulting conditional distribution is:
\begin{equation}
q_\phi\left(z_t \mid x_t, \{C^i_{refine}\}_{i=0}^{2}\right) = \mathcal{N}\left(\mu_t, \mathrm{diag}\left(\exp(\log \mathrm{var}_t)\right)\right).
\end{equation}
During training, the latent is sampled using the reparameterization:
\begin{equation}
z_t = \mu_t + \exp\left(\frac{1}{2}\log \mathrm{var}_t\right) \odot \epsilon, \quad \epsilon \sim \mathcal{N}(0, I).
\end{equation}
The sampled latent is provided to the perceptual decoder together with the shared context:
\begin{equation}
\hat{x}^P_t = D_P\left(\{C^i_{refine}\}_{i=0}^{2}, z_t\right).
\end{equation}
This formulation allows the pretrained latent encoder to parameterize a conditional distribution whose deviation from the standard Gaussian prior is controlled by the KL-divergence regularization described below.

\subsection{Perceptual Objective}
\label{sec:4-2}

The first optimization stage jointly considers pixel fidelity, perceptual similarity, structural similarity, and visual realism. The reconstruction loss is defined as:
\begin{equation}
\begin{aligned}
\mathcal{L}_{rec} =& \lambda_{MSE} D_{MSE}(x_t, \hat{x}^P_t) \\
+& \lambda_{LPIPS} D_{LPIPS}(x_t, \hat{x}^P_t) \\
+& \lambda_{DISTS} D_{DISTS}(x_t, \hat{x}^P_t) \\
+& \lambda_{GAN} \mathcal{L}_{GAN}.
\label{eq:perceptual_reconstruction_loss}
\end{aligned}
\end{equation}
The complete objective for this stage is:
\begin{equation}
\mathcal{L}_P = \mathcal{L}_{rec} + \beta D_{KL}\left(q_\phi\left(z_t \mid x_t, \{C^i_{refine}\}_{i=0}^{2}\right) \| \mathcal{N}(0, I)\right),
\end{equation}
where $D_{MSE}$, $D_{LPIPS}$, and $D_{DISTS}$ measure pixel-wise, perceptual, and structural discrepancies, respectively, and $\mathcal{L}_{GAN}$ encourages visually realistic reconstructions. The KL-divergence term regularizes the conditional distribution of the latent toward the standard Gaussian prior, with $\beta$ controlling the strength of the regularization.

Unlike the fidelity-oriented objective, this stage does not include a rate term for the reconstruction latent. The shared context generation module remains fixed, while the perceptual latent encoder and decoder are optimized jointly.

\subsection{Prior-Sampled Decoder Fine-Tuning and Inference}
\label{sec:4-3}
After the first optimization stage converges, we remove the perceptual latent encoder and retain the perceptual decoder. Instead of obtaining the latent from the encoder, we directly sample it from the standard Gaussian prior at each training iteration:
\begin{equation}
z_t \sim \mathcal{N}(0, I).
\label{eq:prior_sampling}
\end{equation}
The sampled input is provided to the decoder together with the shared context, while the context generation module remains fixed. 

During this stage, the decoder is optimized using only the reconstruction loss defined in Eq.~(\ref{eq:perceptual_reconstruction_loss}):
\begin{equation}
\mathcal{L}^{fine}_P = \mathcal{L}_{rec}.
\end{equation}
The KL-divergence term is no longer required because the latent encoder and its conditional distribution are no longer involved in optimization.

At inference, the randomly sampled input is replaced with a fixed zero-valued input:
\begin{equation}
z_t = 0.
\label{eq:zero_input}
\end{equation}
The final perceptual reconstruction is therefore:
\begin{equation}
\hat{x}^P_t = D_P\left(\{C^i_{refine}\}_{i=0}^{2}, 0\right).
\label{eq:zero_input_reconstruction}
\end{equation}
A fixed input is adopted instead of a randomly sampled one to keep the encoder and decoder consistent. 

\subsection{Intra-Frame Reconstruction}

For intra frames, there is no temporal reference. To construct the contextual representation, the input frame is first downsampled to half resolution and encoded using the intra-frame coding process. The reconstructed low-resolution frame is then transformed into a feature representation that serves as the contextual input for high-resolution reconstruction.

The perceptual decoder reconstructs the high-resolution frame conditioned on this contextual representation. This pathway follows the same reconstruction principle as the inter-frame case: a compact contextual representation provides the primary frame information, while the decoder recovers high-resolution details under the perceptual objective.
     
\section{Experiments}
\label{sec:experiments}
\subsection{Experimental Settings}
\label{sec:settings}

\begin{table*}[t]
    \centering
    \caption{BD-rate(\%, $\downarrow$) comparison of fidelity-oriented compression on 4K sequences. The anchor is the H.266/VVC reference model VTM-13.2 under random access (RA) configuration; negative values indicate bitrate savings at the same PSNR. Best results among all methods are highlighted in \textbf{\underline{bold}}. }
    \label{tab:bd_rate_4k}
    \setlength{\tabcolsep}{3pt}
    \begin{tabular}{llc cc ccccc}
        \toprule
        \multirow{3}{*}{Class} &
        \multirow{3}{*}{Sequence} & 
        \multicolumn{3}{c}{Low delay} &
        \multicolumn{5}{c}{Random access} \\
        \cmidrule(lr){3-5} \cmidrule(lr){6-10}
        & &
        \shortstack{DCVC-DC\\CVPR23~\cite{li2023neural}} &
        \shortstack{DCVC-FM\\CVPR24~\cite{li2024neural}} &
        \shortstack{DCVC-RT\\CVPR25~\cite{dcvcrt2025}} &
        \shortstack{VTM-13.2\\Random access} &
        \shortstack{DCVC-B\\TMM25~\cite{sheng2024bi}} &
        \shortstack{BRHVC\\NeurIPS25~\cite{liu2025brhvc}} &
        \shortstack{HR-NVC\\CVPR26~\cite{ye2026high}} & 
        \shortstack{HR-NVC++\\ OURS}\\
        \midrule
        \multirow{17}{*}{UVG-4K} & Beauty & 3.38 & 26.30 & 20.34 & 0.00 & \textbf{\underline{-6.41}} & 11.76 & 15.69 & 6.15 \\
        & Bosphorus & 20.17 & 13.51 & 9.97 & 0.00 & 4.33 & -2.00 & -11.54 & \textbf{\underline{-19.66}} \\
        & CityAlley & 14.06 & 19.53 & 34.61 & 0.00 & 5.84 & 12.56 & -6.09 & \textbf{\underline{-14.16}} \\
        & FlowerFocus & 28.21 & 75.59 & 15.39 & 0.00 & 107.73 & 89.22 & 15.05 & \textbf{\underline{-25.65}} \\
        & FlowerKids & -0.62 & 15.85 & 60.68 & 0.00 & 35.85 & 19.17 & -18.35 & \textbf{\underline{-31.54}} \\
        & FlowerPan & 15.12 & 46.03 & 70.32 & 0.00 & -18.20 & -14.82 & -30.90 & \textbf{\underline{-33.13}} \\
        & HoneyBee & 2.11 & 14.18 & 12.69 & 0.00 & -7.24 & 3.62 & -19.28 & \textbf{\underline{-26.62}} \\
        & Jockey & 110.04 & 399.04 & 219.23 & \textbf{\underline{0.00}} & 269.03 & 247.32 & 82.87 & 21.47 \\
        & Lips & 13.55 & 55.81 & 40.03 & 0.00 & \textbf{\underline{-6.09}} & 16.86 & 16.17 & 7.75 \\
        & SunBath & -13.28 & 4.10 & 1.78 & 0.00 & 11.24 & 16.37 & 1.41 & \textbf{\underline{-26.06}} \\
        & RaceNight & \textbf{\underline{-11.03}} & -2.47 & 14.78 & 0.00 & 35.88 & 37.75 & 13.16 & -6.80 \\
        & ReadySetGo & 5.65 & 15.02 & 59.95 & 0.00 & 58.99 & 47.22 & 1.49 & \textbf{\underline{-16.17}} \\
        & RiverBank & 20.33 & 22.64 & 33.94 & 0.00 & 1.72 & -0.32 & 2.16 & \textbf{\underline{-1.79}} \\
        & ShakeNDry & -4.69 & 3.85 & -0.28 & 0.00 & \textbf{\underline{-5.27}} & -1.09 & 1.29 & -0.93 \\
        & Twilight & -7.40 & 2.31 & 9.79 & 0.00 & -12.08 & -3.45 & -20.15 & \textbf{\underline{-28.84}} \\
        & YachtRide & -2.27 & 1.11 & 0.91 & 0.00 & 10.40 & 6.61 & 9.87 & \textbf{\underline{-3.07}} \\
        & Average & 12.08 & 44.53 & 37.76 & 0.00 & 30.36 & 30.42 & 3.30 & \textbf{\underline{-12.44}} \\
        \midrule
        \multirow{4}{*}{\shortstack[l]{JVET\\ Class A1}} & Tango2 & -17.92 & -12.14 & -21.24 & 0.00 & 6.04 & 5.39 & -3.22 & \textbf{\underline{-29.93}} \\
        & Campfire & 37.30 & 78.19 & 87.17 & 0.00 & 8.97 & 16.49 & -0.39 & \textbf{\underline{-16.21}} \\
        & FoodMarket4 & -22.90 & -10.18 & -25.92 & 0.00 & 1.07 & 5.17 & -8.17 & \textbf{\underline{-28.38}} \\
        & Average & -1.17 & 18.62 & 13.33 & 0.00 & 5.36 & 9.02 & -3.92 & \textbf{\underline{-24.84}} \\
        \midrule
        \multirow{4}{*}{\shortstack[l]{JVET\\ Class A2}} & ParkRunning3 & 68.07 & 137.57 & 257.47 & \textbf{\underline{0.00}} & 69.87 & 50.91 & 31.86 & 15.68 \\
        & DaylightRoad2 & 22.34 & 38.74 & 73.88 & \textbf{\underline{0.00}} & 60.46 & 42.10 & 37.91 & 0.87 \\
        & CatRobot1 & -16.69 & 2.34 & 16.04 & 0.00 & -18.13 & -17.54 & -17.92 & \textbf{\underline{-29.15}} \\
        & Average & 24.57 & 59.55 & 115.79 & 0.00 & 37.40 & 25.16 & 17.28 & \textbf{\underline{-4.20}} \\
        \midrule
        \multicolumn{2}{c}{Overall Average}
        & 11.98 & 43.04 & 45.07 & 0.00 & 27.91 & 26.79 & 4.22 & \textbf{\underline{-13.01}} \\
        \bottomrule
    \end{tabular}
\end{table*}


\begin{table*}[t]
    \centering
    \begin{threeparttable}
        \caption{BD-rate(\%, $\downarrow$) comparison of fidelity-oriented compression on 1080P sequences. The anchor is the H.266/VVC reference model VTM-13.2 under random access (RA) configuration. Best results among all methods are highlighted in \textbf{\underline{bold}}.}
        \label{tab:bd_rate_1080p}
        \setlength{\tabcolsep}{2.5pt}
        \footnotesize
        \begin{tabular}{@{}l ccc cccccc@{}}
            \toprule
            \multirow{3}{*}{Class} &
            \multicolumn{3}{c}{Low delay} &
            \multicolumn{6}{c}{Random access} \\
            \cmidrule(lr){2-4} \cmidrule(lr){5-10}
            &
            \shortstack{DCVC-DC\\CVPR23~\cite{li2023neural}} &
            \shortstack{DCVC-FM\\CVPR24~\cite{li2024neural}} &
            \shortstack{DCVC-RT\\CVPR25~\cite{dcvcrt2025}} &
            \shortstack{VTM-13.2\\Random access} &
            \shortstack{DCVC-B\\TMM25~\cite{sheng2024bi}} &
            \shortstack{BiECVC\\ACMMM25~\cite{jiang2025biecvc}} &
            \shortstack{BRHVC\\NeurIPS25~\cite{liu2025brhvc}} &
            \shortstack{HR-NVC\\CVPR26~\cite{ye2026high}} &
            \shortstack{HR-NVC++\\OURS} \\
            \midrule

            JCT-VC Class B
            & 24.46 & 28.62 & 56.38
            & 0.00 & 13.74
            & -7.00$^{*}$
            & 7.02
            & 9.80
            & \textbf{\underline{-7.80}} \\

            UVG-1080P
            & 6.06 & 16.98 & 35.58
            & 0.00 & 20.03
            & -9.20$^{*}$
            & 10.86
            & 6.84
            & \textbf{\underline{-11.62}} \\

            MCL-JCV
            & 19.95 & 31.44 & 52.97
            & 0.00 & 23.02
            & 1.40$^{*}$
            & 11.81
            & 14.99
            & \textbf{\underline{-11.92}} \\

            Overall Average
            & 16.82 & 25.68 & 48.31
            & 0.00 & 18.93
            & -4.93$^{*}$
            & 9.90
            & 10.54
            & \textbf{\underline{-10.01}} \\
            \bottomrule
        \end{tabular}

        \begin{tablenotes}[flushleft]
            \footnotesize
            \item[$^{*}$] Results are directly quoted from ~\cite{jiang2026bicrvc} , where the anchor is VTM-23.0 under random access configuration. 
        \end{tablenotes}
    \end{threeparttable}
\end{table*}

\begin{figure*}[tb]
  \centering
  
    \includegraphics[width=\linewidth]{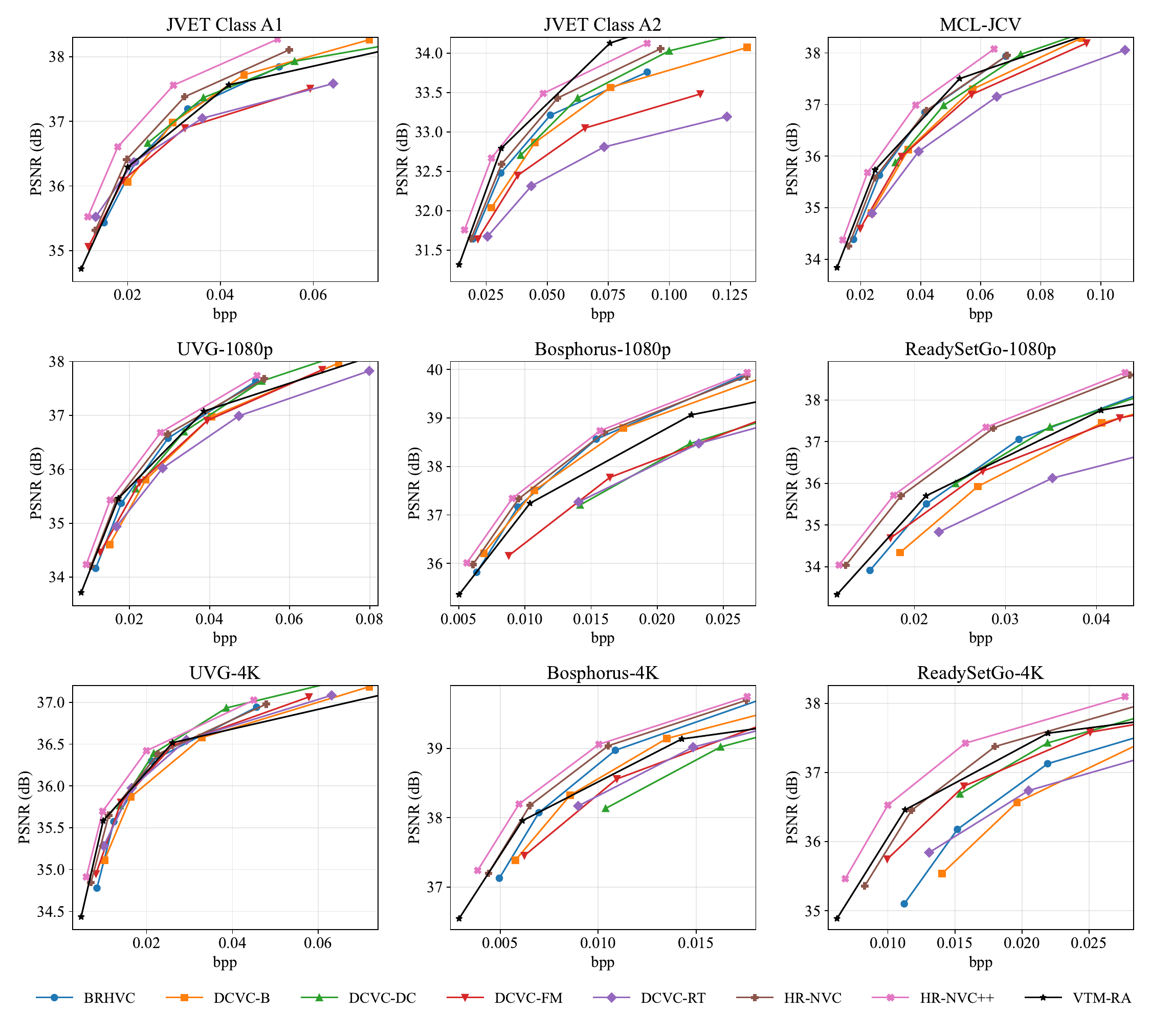}
  \caption{Rate-distortion curves of fidelity-oriented compression on high-resolution benchmarks. Top row: JVET Class A1 (4K), JVET Class A2 (4K), and MCL-JCV (1080P). Middle and bottom rows: Class UVG, Sequence Bosphorus and ReadySetGo, where the same content is coded at 1080P and 4K, respectively. HR-NVC++ consistently improves upon HR-NVC across all benchmarks, and the gain becomes more pronounced when the same content is coded at 4K rather than 1080P, corroborating that the enriched spatial-temporal context is more valuable at higher resolutions.}
  \label{fig:rd_curves1}
\end{figure*}

  

\textbf{Datasets.} 
Our model is first pretrained on the Vimeo90K~\cite{xue2019video} dataset using 7-frame sequences and then fine-tuned on longer video clips collected from raw Vimeo videos, following the multi-stage training protocol of HR-NVC~\cite{ye2026high}. We evaluate HR-NVC++ on six widely used benchmarks: JCT-VC Class B~\cite{hevcdataset}, UVG-1080P~\cite{uvgdatasets}, MCL-JCV~\cite{mcljcv}, UVG-4K, JVET Class A1 and A2. The first three provide 1080P sequences, while the latter three evaluate 4K coding performance. Together, these benchmarks contain diverse motion patterns, textures, and spatial structures. All test sequences are converted to RGB format, and evaluation is performed on the RGB channels.

\textbf{Compared Methods.} 
For fidelity-oriented compression, we compare HR-NVC++ with the H.266/VVC reference model VTM-13.2 under the random access (RA) configuration, and seven NVCs: the low-delay P-frame codecs DCVC-DC~\cite{li2023neural}, DCVC-FM~\cite{li2024neural}, and DCVC-RT~\cite{dcvcrt2025}, and the B-frame codecs DCVC-B~\cite{sheng2024bi}, BiECVC~\cite{jiang2025biecvc}, BRHVC~\cite{liu2025brhvc}, and HR-NVC~\cite{ye2026high}. Note that the BiECVC results are directly quoted from~\cite{jiang2026bicrvc}, where the anchor is VTM-23.0 under the RA configuration. For perceptual-quality-oriented compression, we compare HR-NVC++ with DCVC-FM, DCVC-RT, and PNVC-CR~\cite{Liang2026Perceptual}.

\textbf{Evaluation Metrics.} 
For fidelity-oriented compression, we report PSNR-based BD-rate relative to VTM-13.2 under the RA configuration; a negative BD-rate indicates bitrate savings at matched PSNR. For perceptual-quality-oriented compression, we report LPIPS-VGG~\cite{zhang2018lpips} and DISTS~\cite{ding2022dists}, for which lower values indicate better perceptual similarity, together with the corresponding BD-metric gains using DCVC-FM as the anchor, where positive values indicate perceptual quality improvements at matched bitrate. We further provide qualitative comparisons to examine visual quality and temporal consistency.

\textbf{Evaluation Protocol.} 
We adopt a GOP size of 32 and an intra-period of 32, compressing the first 97 frames of each sequence. 
At relatively high bitrates, sufficient information is available to preserve fine-grained pixel correspondence, making PSNR-based fidelity an appropriate measure. As the bitrate becomes lower, however, strict pixel-level fidelity becomes increasingly difficult to maintain, while perceptual metrics become more relevant to the visual quality of the reconstructed video. We therefore evaluate fidelity-oriented and perceptual-quality-oriented compression over different bitrate ranges. 

For the compared methods with a broad bitrate range, including DCVC-FM and DCVC-RT, we evaluate eight approximately uniformly distributed bitrate points, where the four higher-bitrate points are used for fidelity evaluation and the four lower-bitrate points for perceptual evaluation. For methods with a more limited bitrate range, we evaluate four approximately uniformly distributed bitrate points.
For HR-NVC++, fidelity-oriented models are first trained with the MSE objective, covering the higher-bitrate range. Perceptual model is then adapted as described in Section~\ref{sec:perceptual}; since the frame-dependent latent is no longer transmitted, the resulting operating points naturally shift to the lower-bitrate range. 

\subsection{Fidelity-oriented Compression}
\label{sec:fidelity_results}

\begin{figure*}[tb]
  \centering
  
    \includegraphics[width=\linewidth]{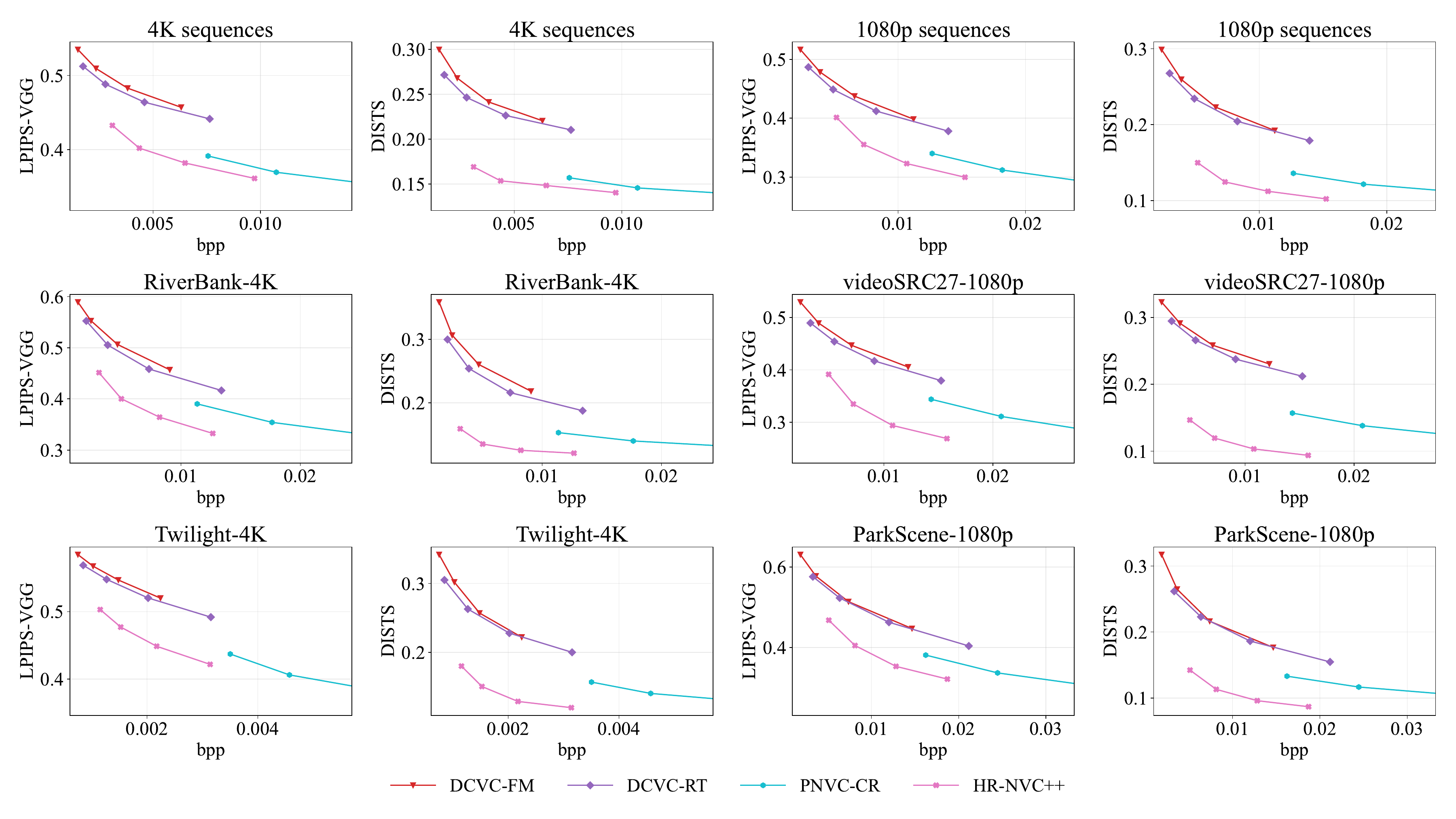}
  \caption{Perceptual quality-rate curves of the perceptual-quality-oriented compression on high-resolution benchmarks. The first two columns report LPIPS-VGG and DISTS on 4K sequences, and the last two columns on 1080P sequences; lower values indicate better perceptual similarity. HR-NVC++ achieves the best perceptual similarity across both metrics and both resolutions.}
  \label{fig:perceptual_curves}
\end{figure*}

\begin{table*}[t]
    \centering
    \caption{BD-metric ($\uparrow$) comparison of perceptual-quality-oriented compression on high-resolution benchmarks. The anchor is DCVC-FM. Positive values indicate perceptual quality gains (lower LPIPS-VGG/DISTS) at the same bitrate. Best results among all methods are highlighted in \textbf{\underline{bold}}.}
    \label{tab:perceptual_metric}
    \begin{tabular}{cc ccc ccc c}
        \toprule
        \multirow{2}{*}{Metric} &
        \multirow{2}{*}{Method} &
        \multicolumn{3}{c}{4K} &
        \multicolumn{3}{c}{1080P} &
        \multirow{2}{*}{Average} \\
        \cmidrule(lr){3-5} \cmidrule(lr){6-8}
        & &
        {UVG-4K} &
        {JVET Class A1} &
        {JVET Class A2} &
        {JCT-VC Class B} &
        {UVG-1080P} &
        {MCL-JCV} & \\
        \midrule
        \multirow{3}{*}{LPIPS-VGG} & 
        DCVC-FM~\cite{li2024neural} 
        & 0.00 & 0.00 & 0.00 & 0.00 & 0.00 & 0.00
        & 0.00 \\
        & DCVC-RT~\cite{dcvcrt2025} 
        & 0.0057 & 0.0102 & 0.0039 & 0.0084 & 0.0036 & 0.0051
        & 0.0062 \\
        & HR-NVC++ 
        & \textbf{\underline{0.0771}} & \textbf{\underline{0.0640}} & \textbf{\underline{0.0611}} & \textbf{\underline{0.0861}} & \textbf{\underline{0.0917}} & \textbf{\underline{0.0695}} & \textbf{\underline{0.0749}} \\
        \midrule
        \multirow{3}{*}{DISTS} & 
        DCVC-FM~\cite{li2024neural} 
        & 0.00 & 0.00 & 0.00 & 0.00 & 0.00 & 0.00 
        & 0.00 \\
        & DCVC-RT~\cite{dcvcrt2025} 
        & 0.0060 & 0.0104 & 0.0023 & 0.0044 & 0.0050 & 0.0035
        & 0.0053 \\
        & HR-NVC++ 
        & \textbf{\underline{0.0844}} & \textbf{\underline{0.0858}} & \textbf{\underline{0.0546}} & \textbf{\underline{0.0754}} & \textbf{\underline{0.0943}} & \textbf{\underline{0.0930}} & \textbf{\underline{0.0813}} \\
        \bottomrule
    \end{tabular}
\end{table*}

We first evaluate the fidelity-oriented compression of HR-NVC++. Tables~\ref{tab:bd_rate_4k} and~\ref{tab:bd_rate_1080p} summarize the PSNR-based BD-rate results on the 4K and 1080P benchmarks, respectively. With VTM-13.2 under the RA configuration as the anchor, HR-NVC++ achieves average BD-rate savings of 13.01\% on the 4K benchmarks and 10.01\% on the 1080P benchmarks, outperforming all compared NVCs. 

Averaged over all six benchmarks, HR-NVC++ achieves a BD-rate of -11.51\%, compared with 7.38\% for HR-NVC. The gain over HR-NVC is particularly informative. Based on HR-NVC, HR-NVC++ adds complementary spatial information and confidence-guided spatial-temporal refinement. The consistent reduction in BD-rate therefore indicates that enriching the contextual representation beyond bi-directional temporal reference modeling provides additional benefits for high-resolution compression.

Compared with the other neural P-frame and B-frame codecs, HR-NVC++ achieves the best average BD-rate on every benchmark. It is worth noting that the Jockey sequence is an outlier on which several codecs degrade sharply, with BD-rates exceeding 100\% for some methods; such extreme values can distort the average BD-rate. Excluding this sequence, HR-NVC++ still achieves the best average BD-rate on every benchmark and the best per-sequence result on the majority of sequences.

The rate-distortion curves in Fig.~\ref{fig:rd_curves1} further show that the gain becomes more pronounced when the same content is coded at 4K rather than 1080P, supporting the motivation that spatial information becomes increasingly valuable when temporal correspondence is less reliable at high spatial resolutions. 
Overall, these results indicate that the proposed confidence-guided spatial-temporal context provides more useful predictive information than relying on bi-directional temporal references alone, translating the richer contextual representation into consistent coding gains across both 1080P and 4K content.

\begin{figure*}[tb]
  \centering
  
    \includegraphics[width=0.98\linewidth]{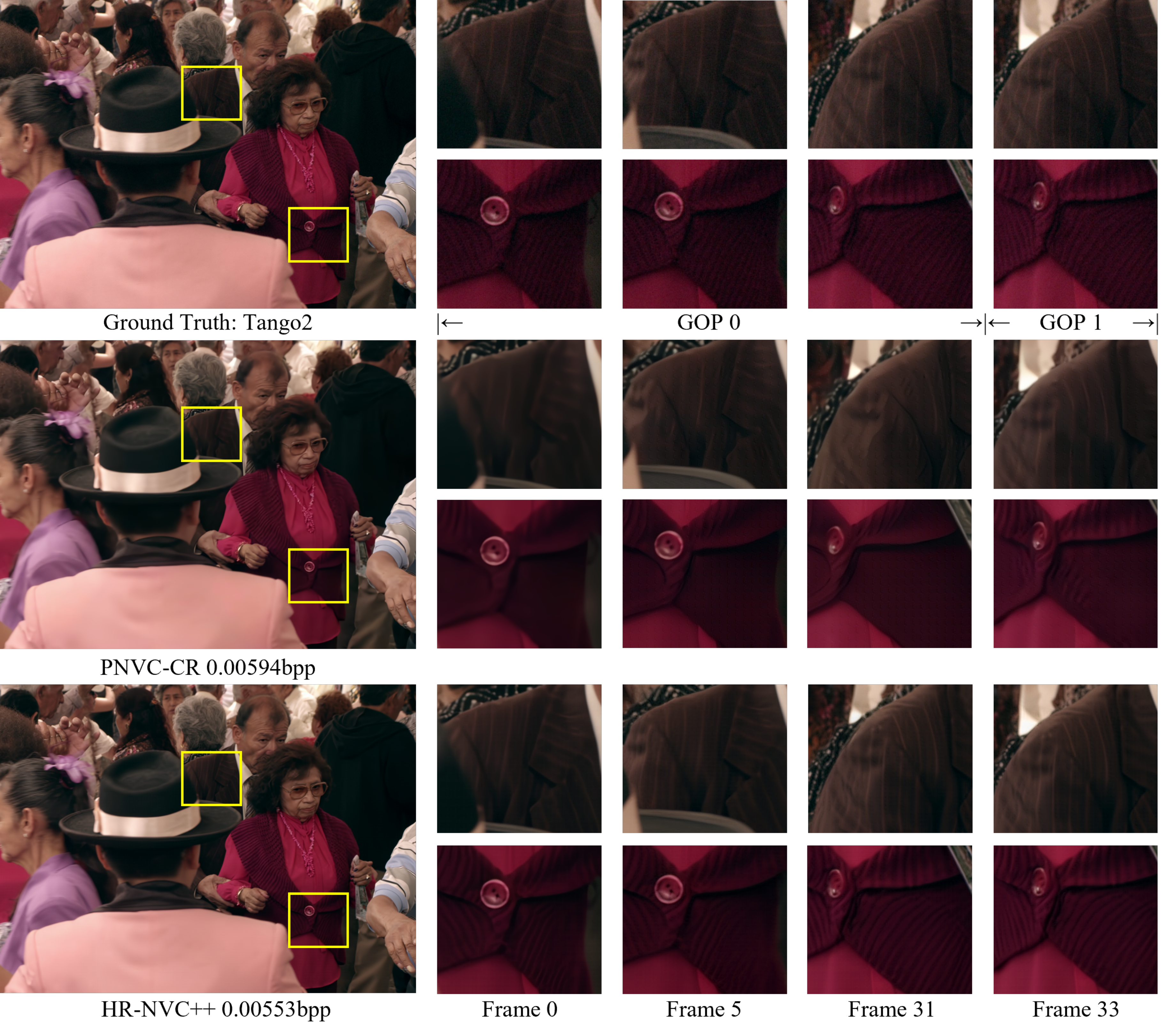}
  \caption{Temporal consistency on Tango2 (JVET Class A1) across GOPs. The enlarged regions cover the stripes on the suit jacket and the texture of the sweater of frames 0, 5, and 31 (GOP 0) and frame 33 (GOP 1). HR-NVC++ preserves structures and keeps them stable across the GOP. In contrast, PNVC-CR nearly loses the sweater texture in GOP 0, partially recovers it in GOP 1, and produces stripes that alternate between blurred and sharp from frame to frame.}
  \label{fig:perceptual_temporal_a}
\end{figure*}

\begin{figure*}[tb]
  \centering
  
    \includegraphics[width=0.98\linewidth]{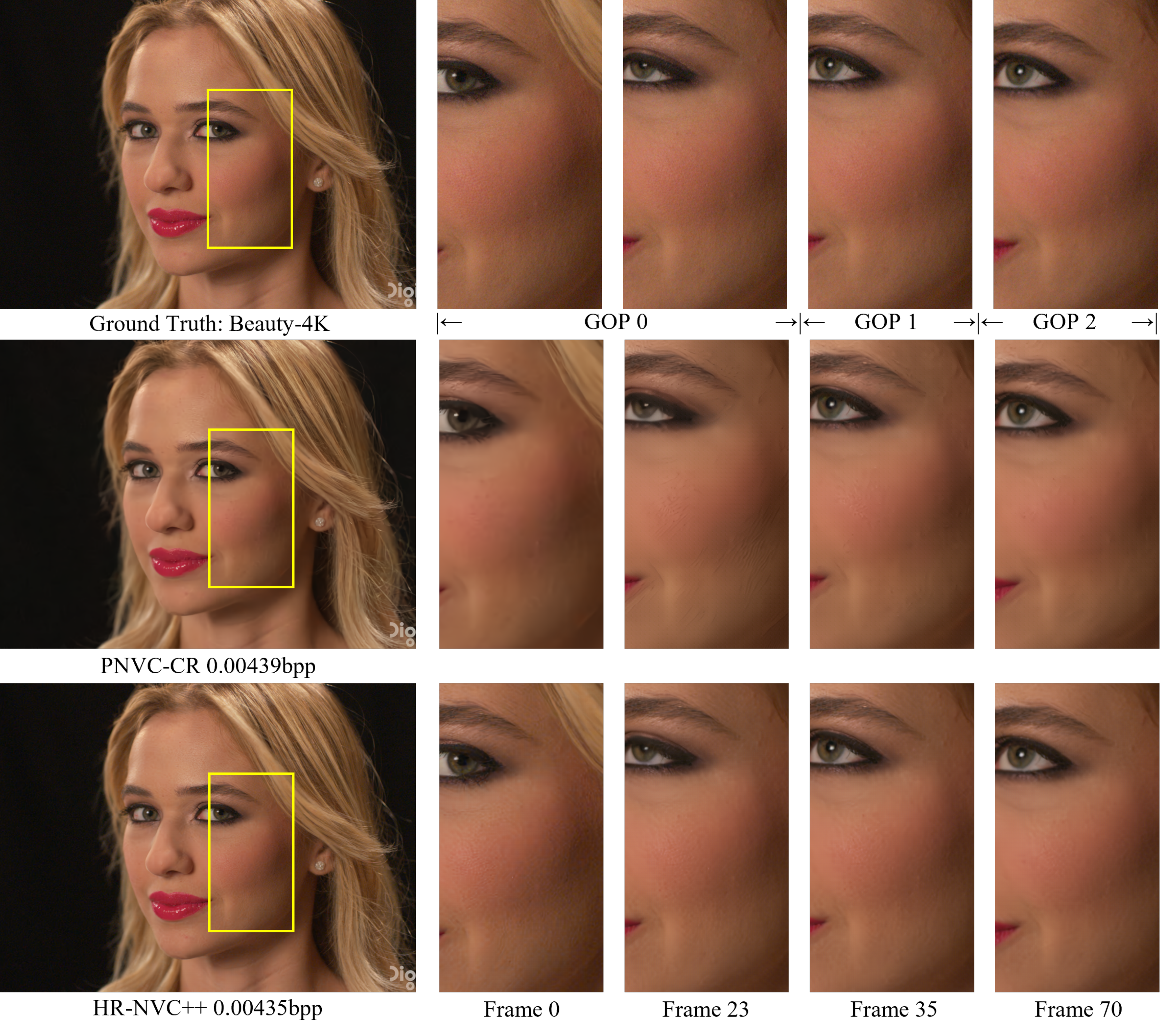}
  \caption{Temporal consistency on Beauty-4K (UVG-4K) across GOPs. The enlarged region covers the facial area of frames 0 and 23 (GOP 0), frame 35 (GOP 1), and frame 70 (GOP 2). HR-NVC++ retains the facial texture throughout all sampled frames, whereas PNVC-CR fails to preserve these details and even produces spurious texture patterns at frame 23, indicating that frame-level perceptual gains do not necessarily translate into temporally coherent video quality.}
  \label{fig:perceptual_temporal_b}
\end{figure*}

\begin{figure*}[tb]
  \centering
  
    \includegraphics[width=\linewidth]{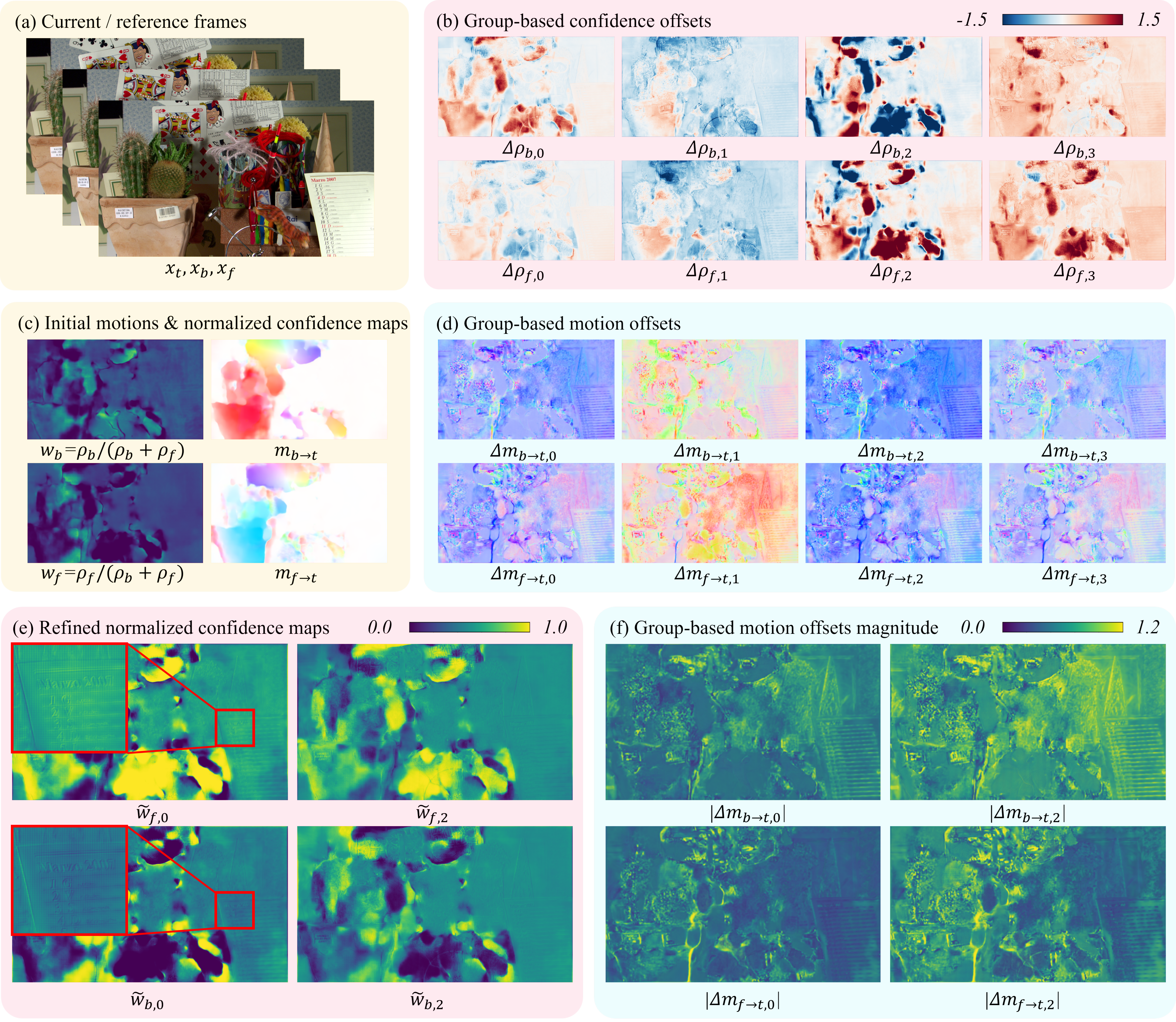}
  \caption{Visualization of the confidence-guided spatial-temporal context construction on the Cactus sequence (JCT-VC Class B). (a) The current frame $x_t$ is frame 16 and the two reference frames $\hat{x}_f$ and $\hat{x}_b$ are frames 0 and 32, respectively. (b) The predicted group-wise confidence offsets $\Delta\rho_{f,g}$ and $\Delta\rho_{b,g}$. (c) The initial bi-directional motion fields $m_{f\rightarrow t}$ and $m_{b\rightarrow t}$, together with the normalized confidence weights $w_f=\rho_f/(\rho_f+\rho_b)$ and $w_b=\rho_b/(\rho_f+\rho_b)$. (d) The predicted group-wise motion offsets $\Delta m_{f\rightarrow t,g}$ and $\Delta m_{b\rightarrow t,g}$. (e) The refined normalized confidence weights $\tilde{w}_{f,g}$ and $\tilde{w}_{b,g}$. (f) The magnitudes of the group-wise motion offsets $|\Delta m_{f\rightarrow t,g}|$ and $|\Delta m_{b\rightarrow t,g}|$. The subscript $g$ denotes the group index; eight groups are predicted in total, and only part of them are shown. All visualized maps are at the highest-resolution scale of the hierarchical context. }
  \label{fig:ablation3}
\end{figure*}

\subsection{Perceptual-quality-oriented Compression}
\label{sec:perceptual_results}

Fig.~\ref{fig:perceptual_curves} plots LPIPS-VGG and DISTS against bitrate for HR-NVC++, DCVC-FM, DCVC-RT, and PNVC-CR over the low-bitrate range on both 4K and 1080P sequences, where lower values indicate better perceptual similarity. HR-NVC++ consistently outperforms all compared methods, both on the dataset averages and on individual sequences at both resolutions. 

For HR-NVC++, the perceptual mode reconstructs each inter frame using the shared context as the sole informative input. Its inter-frame bitstream carries only the motion information and the spatial complement, which substantially reduces the bitrate and enables compression at extremely low bitrates while maintaining favorable perceptual quality. This indicates that the shared context already provides the dominant content required for perceptually favorable compression. In addition, the advantage holds on both 4K and 1080P content, suggesting that the enriched spatial-temporal context benefits perceptual compression across resolutions and bitrate points. 

Table~\ref{tab:perceptual_metric} further quantifies these gains in terms of BD-metric with DCVC-FM as the anchor. HR-NVC++ achieves average BD-metric improvements of 0.0749 (LPIPS-VGG) and 0.0813 (DISTS) over the six benchmarks, substantially larger than those of DCVC-RT (0.0062 and 0.0053).


Frame-wise metrics such as LPIPS and DISTS, however, cannot capture temporal consistency: improving frame-level perceptual quality can potentially introduce temporal inconsistency across consecutive frames. We therefore further examine the perceptual reconstruction over consecutive frames. Figs.~\ref{fig:perceptual_temporal_a} and~\ref{fig:perceptual_temporal_b} present frames sampled across consecutive GOPs from Tango2 and Beauty-4K, respectively, comparing HR-NVC++ with PNVC-CR at similar bitrates.

On Tango2, HR-NVC++ preserves both the stripes on the suit jacket and the texture of the sweater, and these fine details remain stable across the GOP. In contrast, PNVC-CR nearly loses the sweater texture in GOP 0, while part of the texture reappears in GOP 1; the stripes on the suit jacket likewise alternate between blurred and sharp from frame to frame.

On Beauty-4K, HR-NVC++ retains the facial texture throughout the sampled frames, whereas PNVC-CR fails to preserve these details and even produces spurious texture patterns at frame 23. Such unstable reconstructions indicate that the frame-level perceptual gains of PNVC-CR do not translate into temporally coherent video quality. By comparison, the stable object boundaries, textures, and local structures produced by HR-NVC++ across consecutive frames and GOPs demonstrate that the perceptual compression benefits from the spatial-temporal context without sacrificing temporal coherence, complementing the frame-level evaluation above.

\subsection{Ablation Study}
\textbf{Effectiveness of Spatial-Temporal Context.} 
We first analyze the impact of motion resolution and the spatial complement on high-resolution coding. HR-NVC-hm is a variant of our previous HR-NVC~\cite{ye2026high} in which motion estimation and motion coding are performed at half of the original resolution, with all other components unchanged. HR-NVC-hm++ further equips this variant with the proposed confidence-guided spatial-temporal fusion, but replaces the spatial complement with a zero latent that is not coded; HR-NVC++ additionally codes the spatial complement, while sharing the same half-resolution motion modeling as HR-NVC-hm++.

To intuitively understand what the proposed context learns, Fig.~\ref{fig:ablation3} visualizes the bi-directional motion fields, the initial and refined confidence maps, and the group-wise motion and confidence offsets on the Cactus sequence. The refined confidence maps suppress the unreliable direction in occluded and newly exposed regions, while the motion offsets concentrate around moving-object boundaries where the initial estimates are most error-prone, indicating that the refinement exploits the bi-directional temporal cues in a content-adaptive manner.

\begin{figure}[tb]
  \centering
  
    \includegraphics[width=\linewidth]{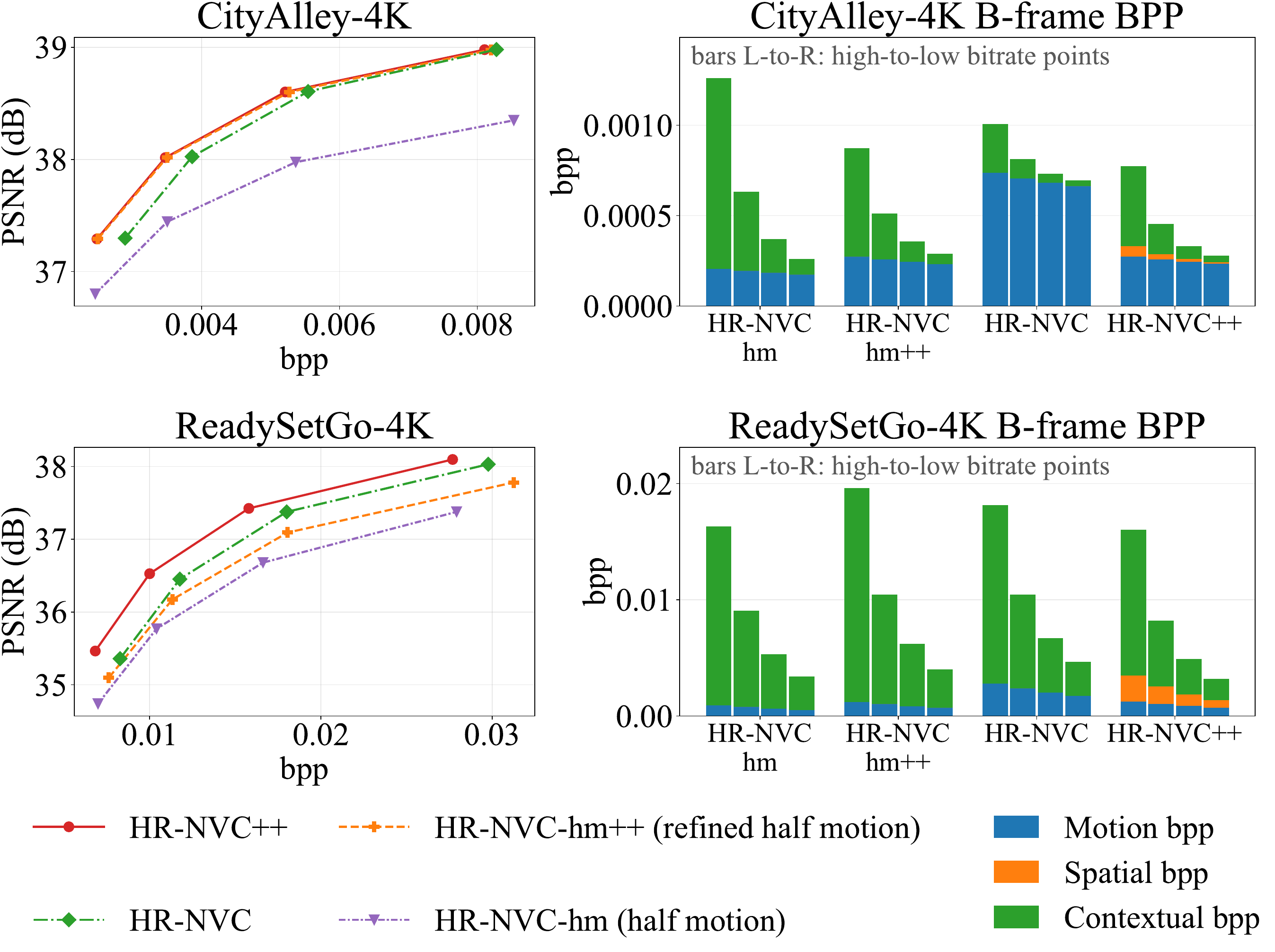}
  \caption{Ablation of the proposed spatial-temporal context on CityAlley-4K and ReadySetGo-4K (UVG-4K). Four variants: HR-NVC, HR-NVC-hm (half-resolution motion estimation and coding), HR-NVC-hm++ (half-resolution motion with the confidence-guided fusion, but with the spatial complement replaced by zero latent), and HR-NVC++. }
  \label{fig:ablation1}
\end{figure}

Fig.~\ref{fig:ablation1} compares the four variants on CityAlley-4K and ReadySetGo-4K, together with the average B-frame bitrate decomposition into motion, spatial, and contextual components. The two sequences reveal complementary behaviors. On the static sequence CityAlley-4K, simply reducing the motion resolution (HR-NVC-hm) significantly degrades the performance compared with HR-NVC. Nevertheless, equipping half-resolution motion with the proposed fusion structure (HR-NVC-hm++) already surpasses HR-NVC with full-resolution motion and nearly matches HR-NVC++. Correspondingly, the bitrate decomposition shows that the spatial branch accounts for only a negligible share of the total bitrate in HR-NVC++. This indicates that, for static content, the confidence-guided group-wise fusion itself is sufficient, and the spatial complement contributes little because temporal references already explain the scene well.

The dynamic sequence ReadySetGo-4K exhibits the opposite behavior. Reducing the motion resolution (HR-NVC-hm) causes a clear PSNR degradation, and the fusion structure alone (HR-NVC-hm++) recovers only part of the loss. The gap is closed only when the spatial complement is actually coded (HR-NVC++), which surpasses HR-NVC with full-resolution motion while operating on cheaper half-resolution motion. Consistent with this, the spatial branch occupies a substantial share of the bitrate on this sequence in HR-NVC++. This shows that, under complex motion where temporal correspondence is unreliable, the spatial complement carries genuinely complementary information that cannot be recovered by temporal modeling alone, and is therefore necessary.

Overall, these results support the design motivation of HR-NVC++: the confidence-guided fusion exploits temporal evidence where it is reliable, while the spatial complement is allocated bits where temporal prediction fails, and the bitrate allocation between the two sources adapts to the content automatically.

\begin{figure}[tb]
  \centering
  
    \includegraphics[width=\linewidth]{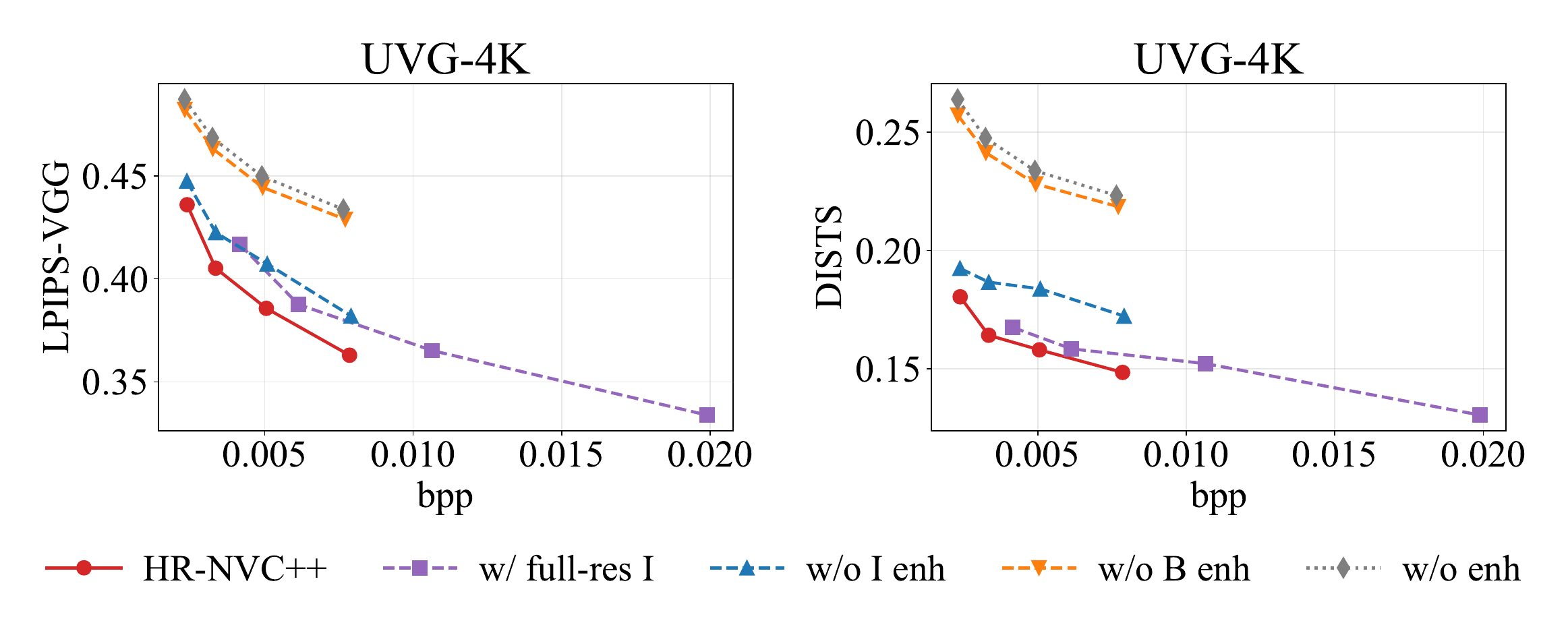}
  \caption{Ablation of the perceptual enhancement on UVG-4K. Five variants: without enhancement on both intra and B frames (w/o enh), with enhancement on intra frames only (w/o B enh), on B frames only (w/o I enh), on both frame types (HR-NVC++), and a variant that directly codes intra frames at full resolution (w/ full-res I).}
  \label{fig:ablation2}
\end{figure}

\textbf{Effectiveness of Perceptual Enhancement.} 
We next ablate the perceptual-quality-oriented compression mode on UVG-4K in terms of LPIPS-VGG and DISTS, as shown in Fig.~\ref{fig:ablation2}. In HR-NVC++, an intra frame is coded at half resolution, and the perceptual decoder then reconstructs the full-resolution frame using the context extracted from the reconstructed low-resolution intra frame. We compare four enhancement configurations: without enhancement on both intra and B frames (w/o enh), with enhancement on intra frames only (w/o B enh), with enhancement on B frames only (w/o I enh), and with enhancement on both (HR-NVC++). In addition, we evaluate a variant that directly codes intra frames at full resolution (w/ full-res I).

Compared with full-resolution intra coding (w/ full-res I), coding intra frames at half resolution without enhancement (w/o I enh) substantially reduces the bitrate but suffers a marked degradation in DISTS. By enhancing the frames reconstructed from the compact half-resolution intra context, HR-NVC++ achieves better perceptual quality than full-resolution intra coding while retaining the low bitrate. This indicates that the intra-frame enhancement cannot be replaced by simply allocating more bits to intra frames: it recovers the details sacrificed by compact half-resolution coding, which would otherwise require a considerable number of bits to preserve explicitly.

We then analyze the contribution of the enhancement on each frame type. Removing the enhancement on both frame types (w/o enh) leads to the worst perceptual similarity, and enhancing either frame type alone brings only limited improvement. The best results are obtained when both frame types are enhanced (HR-NVC++), indicating that enhanced intra frames also benefit the subsequent B frames by providing higher-quality references for context generation.

\subsection{Computational Complexity}
\label{sec:complexity}
Table~\ref{tab:complexity} compares HR-NVC++ with existing neural B-frame codecs in terms of model size and computational complexity. The parameter counts cover the complete codec, including both the intra-frame and B-frame coding models. Although HR-NVC++ introduces additional modules for spatial coding and spatial-temporal context construction, which increase the parameter count (80.96M vs. 54.03M of HR-NVC), its per-pixel complexity is the lowest among all compared codecs: 2518 kMACs/pixel, vs. 2646 for HR-NVC, 2941 for DCVC-B, and 3917 for BRHVC.

\begin{figure}[t]
  \centering
  
    \includegraphics[width=\linewidth]{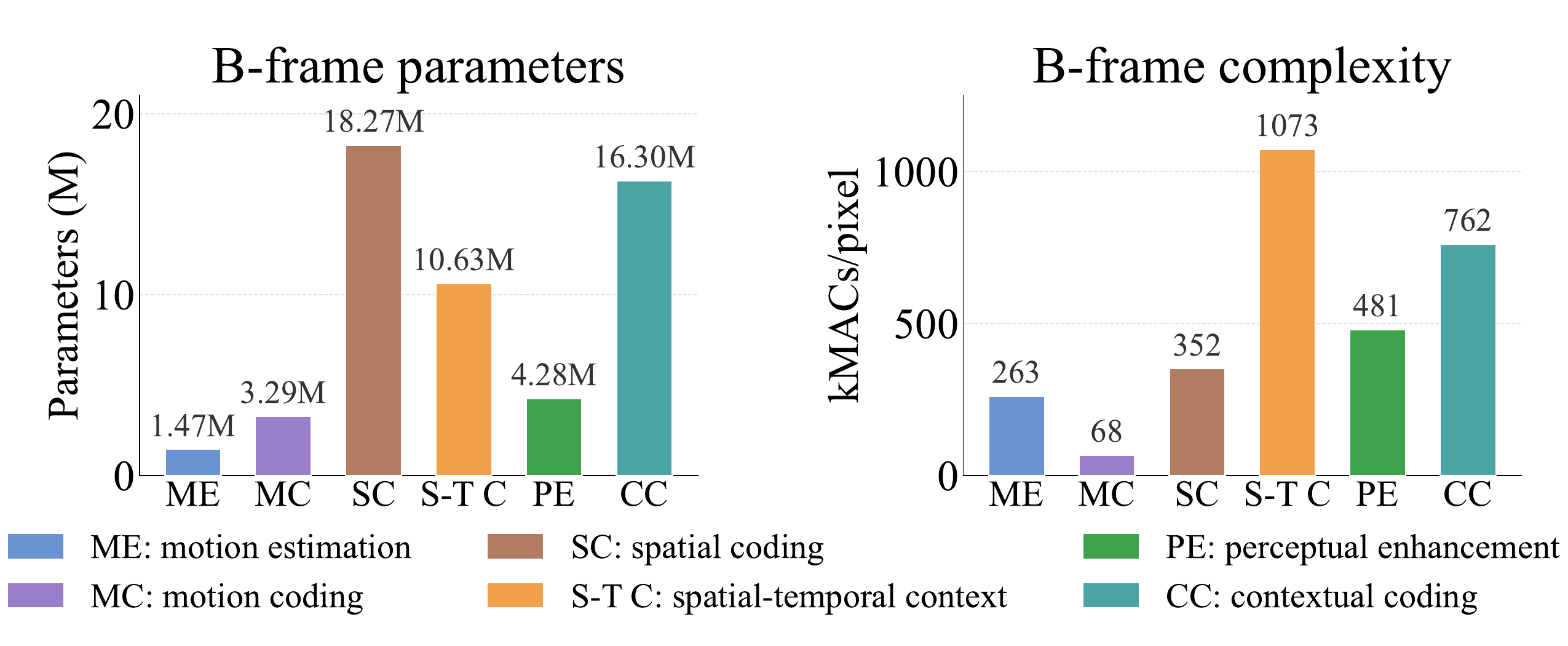}
  \caption{Module-wise breakdown of B-frame parameters and computational complexity. The fidelity mode of HR-NVC++ comprises ME, MC, SC, S-T C, and CC (2518 kMACs/pixel), while the perceptual mode replaces CC with PE (2237 kMACs/pixel).}
  \label{fig:complexity}
\end{figure}

The module-wise breakdown of the B-frame coding modules is shown in Fig.~\ref{fig:complexity}. The efficiency mainly comes from the half-resolution motion modeling: motion estimation and motion coding together account for only 331 kMACs/pixel, about 13\% of the total, and their parameter cost is also marginal. The most computationally intensive module is the spatial-temporal context construction (1073 kMACs/pixel), which operates on multi-scale features, while the spatial coding branch carries the largest parameter count (18.27M) due to its entropy model. 

\begin{table}[t]
    \centering
    \caption{Computational complexity comparison. The parameter counts cover the complete codec.}
    \label{tab:complexity}
    \begin{tabular}{lcc}
        \toprule
        Method & Params (M) & kMACs/pixel  \\
        \midrule
        DCVC-B~\cite{sheng2024bi} & 55.38 & 2941 \\  
        BRHVC~\cite{liu2025brhvc} & 60.43 & 3917 \\
        HR-NVC~\cite{ye2026high} & 54.03 & 2646 \\
        \textbf{HR-NVC++ (fidelity)} & 80.96 & 2518 \\
        \textbf{HR-NVC++ (perceptual)} & 68.94 & 2237 \\
        \bottomrule
    \end{tabular}
\end{table}

The perceptual mode further reduces the complexity. Since the frame-dependent latent is no longer coded, the contextual coding module (762 kMACs/pixel) is removed and replaced by the perceptual enhancement module (481 kMACs/pixel), yielding 2237 kMACs/pixel in total, together with a reduced parameter count (68.94M vs. 80.96M). The perceptual mode is therefore cheaper to run than the fidelity mode.

Overall, HR-NVC++ improves the rate-distortion performance over HR-NVC while reducing the computational complexity. These results indicate that the proposed spatial coding and spatial-temporal context construction improve coding efficiency without increasing the per-pixel computational burden.    
\section{Conclusion}
\label{sec:conclusion}
In this work, we present HR-NVC++, an ultra-high-resolution NVC approach that unifies fidelity-oriented and perceptual-quality-oriented compression with a shared contextual representation. By constructing a confidence-guided spatial-temporal context that jointly exploits bi-directional temporal cues, their correspondence reliability, and complementary spatial information, HR-NVC++ supports both reconstruction objectives within a single framework: the fidelity mode retains explicit latent coding for accurate pixel-level reconstruction, while the perceptual mode reconstructs directly under the shared context. Extensive experiments on multiple 1080P and 4K benchmarks demonstrate that HR-NVC++ achieves state-of-the-art performance in both fidelity and perceptual modes.

{
    \bibliographystyle{IEEEtran}
    \bibliography{main}
}


 





\end{document}